# Large Deviations Sum-queue Optimality of a Radial Sum-rate Monotone Opportunistic Scheduler

Bilal Sadiq and Gustavo de Veciana
Department of Electrical and Computer Engineering
The University of Texas at Austin

*Abstract*—A centralized wireless system is considered that is serving a fixed set of users with time varying channel capacities. An opportunistic scheduling rule in this context selects a user (or users) to serve based on the current channel state and user queues. Unless the user traffic is symmetric and/or the underlying capacity region a polymatroid, little is known concerning how performance optimal schedulers should tradeoff *maximizing current service rate* (being opportunistic) versus *balancing unequal queues* (enhancing user-diversity to enable future high service rate opportunities). By contrast with currently proposed opportunistic schedulers, e.g., MaxWeight and Exp Rule, a radial sum-rate monotonic (RSM) scheduler de-emphasizes queue-balancing in favor of greedily maximizing the system service rate as the queue-lengths are scaled up linearly. In this paper it is shown that an RSM opportunistic scheduler, p-Log Rule, is not only throughput-optimal, but also maximizes the asymptotic exponential decay rate of the sum-queue distribution for a two-queue system. The result complements existing optimality results for opportunistic scheduling and point to RSM schedulers as a good design choice given the need for robustness in wireless systems with both heterogeneity and high degree of uncertainty.

## I. INTRODUCTION

We consider a wireless node shared by two users, where each user's data arrives to a queue at the node as a random process and each user's channel, in terms of data rate that it can support, varies randomly over time. If the channel state is available, a policy can schedule users so as to exploit favorable channels, e.g., schedule the user which currently has the highest rate – this is referred to as opportunistic scheduling (see, e.g., [1], [2]). More generally, a scheduler can be permitted to allocate rates to the users from a compact polytope, where the polytope depends on the channel state.

An opportunistic or channel-aware scheduler, however, may not be stable, i.e., keep the user queues bounded, unless it is chosen carefully, e.g., using prior knowledge of the arrival and channel process. Except in some degenerate cases, in order to ensure stability for all possibly stabilizable arrival processes, an opportunistic scheduler must be both channel- and queue-aware, and must tradeoff *maximizing the current service rate* (scheduling the queue seeing the highest rate) with *balancing unequal queues* (scheduling the longest queue). Note that by balancing queues, one can increase the likelihood of more non-empty queues in the future and, therefore, can potentially achieve higher service rates in the future. Moreover, the *performance* of any scheduler, in terms of Quality-of-Service (QoS) measures like mean delay or probability of queue overflow, depends on how this tradeoff is made.

Much work to date has focused on engineering queue-and-channel-aware schedulers that are *throughput-optimal*, i.e., schedulers that can achieve stability without any knowledge of arrival or channel statistics, if stability is at all feasible. Examples are MaxWeight [3], Exponential (Exp) rule [4], and Log rule [5]. In fact, necessary and sufficient conditions for MaxWeight-type schedulers to be throughput-optimal have been shown in [6], [7].

Stability, however, is a weak form of optimality. In view of various QoS goals, it is of interest to study schedulers that are *delay-optimal*, e.g., schedulers that minimize the overall average delay (per data unit) seen by users, or ones which minimize the probability that either the sum-queue or the longest queue overflows a large buffer. These schedulers are harder to characterize for channels/servers[1] with time-varying capacity, but some results are available that we briefly discuss next.

In [8] and [9] the Longest-Connected-Queue (LCQ) and Longest-Queue-Highest-Possible-Rate (LQHPR) scheduling policies are introduced. Strong results are shown for these policies; they stochastically minimize the max- and sum-queue processes, and thus also the tails of max- and sum-queue distributions and mean delay. However, in addition to assuming certain symmetry conditions on arrival and channel statistics, [8] is limited to on-off channel capacities where only a single queue can be scheduled per time slot, and [9] assumes that the scheduler can allocate service rates from the information theoretic polymatroid multiuser capacity region associated with the current channel state. In both cases, the above-mentioned tradeoff between queue balancing and service rate maximization is absent. Indeed in [8], all policies that pick a connected queue result in the same service rate, whereas, in the case of [9], all policies that pick a service vector from the maximal points of the current capacity region, i.e., points on the max-sum-rate face, result in the same total service rate. Thus one can achieve the *queue balancing* goal, without compromising the total *service rate*. Not surprisingly, in both cases the optimal policy turns out to be greedy, in that it allocates as much service rate as possible to the longest/longer queues.

First draft June 20, 2009; revised December 17, 2009. This research was supported in part by grants NSF CNS-0721532, AFOSR FA9550-07-1-0428, and by Intel. These results were presented in part at the 46th Allerton Conference on Communications, Control, and Computing, September 2008.

[1]The terms channel and server, and sometimes the terms user and queue, are used interchangeably through this section.

A related server allocation problem is studied in [10]. The paper considers minimizing the mean delay in a two queue system where each queue has a dedicated server and a third server can be dynamically shared between them. As a result, the two queues can be allocated service rates from a polymatroid capacity region, thus the objective of queue balancing can again be achieved without compromising the total service rate. However, without the underlying symmetry assumptions of [8] and [9], and using a dynamic programming approach, only the existence of a monotone increasing switching curve on the state space of queue process is shown; note that the switching curve under LCQ and LQHPR policies lies along the line where both queues are equal. For a system with a general compact, convex, and coordinate convex capacity region and any finite number of queues, [11] gives a large deviations principle (LDP) for transient queue process under MaxWeight scheduler. This LDP can be used to compute, e.g, the asymptotic probability of sum-queue or max-queue overflow, as well as the corresponding likely modes of overflow. Although the capacity region is not changing over time, the region is such that a scheduler must tradeoff maximizing total service rate with balancing unequal queues. Therefore this result is insightful in relating the modes overflow to the tradeoff made by the MaxWeight scheduler. A more recent work [12] gives many sources large deviations for MaxWeight scheduler for a similar capacity region.

Finally, relaxing the symmetry assumptions of [8] and [9], the works in [13]–[15] consider the asymptotic probability of max-queue overflow. The server capacity in [13], though time-varying, is identical for all users at any given time, thus the need to tradeoff queue-balancing versus service rate maximization is again absent. In fact, the sum-queue process in [13] is identical for all work conserving schedulers. However [14] and [15] consider a server with *asynchronously* time-varying capacity across users. [14] studies the asymptotic probability of max-queue overflow under MaxWeight scheduler and shows that for a given system, as the exponent of queue length in the MaxWeight scheduler, $\alpha$, becomes large, the asymptotic probability of max-queue overflow under MaxWeight approaches the minimum achievable under any other scheduler. A stronger result is shown in [15], that is, the Exp rule scheduler in fact minimizes the steady state asymptotic probability of max-queue overflow. Indeed the models in [14] and [15] accurately capture a wireless channel shared by heterogenous users, and exhibit the tradeoff between queue-balancing and service rate maximization. Existence of this tradeoff also implies that, unlike the LCQ and LQHPR policies, the asymptotic optimality of Exp rule does not translate to minimizing the asymptotic probability of sum-queue overflow or the mean delay. In order to minimize the asymptotic probability of max-queue overflow, the desired[2] mode of overflow is one where all queues (more precisely, the set of overflowing queues which then exclusively share the server,) grow at the same rate and overflow at the same time. This constrains the system throughput, while, of course, keeping the queue lengths *equal* across users.

*Contributions*: In this paper we address the problem of minimizing the asymptotic probability of sum-queue overflow. First, we give a tight lower bound on this probability under *any* scheduler. By contrast with the desired mode overflow to minimize the asymptotic probability of *max*-queue overflow, we will see that in order to minimize the asymptotic probability of sum-queue overflow, the desired mode of overflow is one where the system throughput is the highest possible and queues may build up at different rates. Second, we show that a radial sum-rate monotone scheduler (see [5] or Section III for definition of radial sum-rate monotonicity), called the pseudo-Log rule, minimizes the asymptotic probability of sum-queue overflow. Although our focus is on overflows of the sum-queue instead of overflows of the max-queue as in [15], the general technique of proof in [15] lends itself well to our problem and we rely heavily on the results developed therein. Other desirable features of radial sum-rate monotone schedulers have been explained in [5], which include,

(a) reducing mean delay;
(b) graceful degradation of service in terms of fraction of users that can meet their Quality-of-Service requirements during transient overloads;
(c) robustness to uncertainty in traffic and channel statistics.

The rest of the paper is organized as follows. The system model is described in Section II. Queue-and-channel aware schedulers of interest, namely, MaxWeight, Exp rule, and Log rule, and the property of radial sum-rate monotonicity are reviewed in Section III, followed by the introduction of pseudo-Log scheduling rule in Section IV. The main result of the paper is summarized in Section V. Some preliminary discussion and relevant large deviation principles follow in Section VI. The proofs for the lower and the upper bounds stated in the main result of the paper are given in Section VII and VIII respectively. After defining local fluid sample paths and developing essential results (summarized in Table I) in Section IX, the optimality of the p-Log rule, i.e., the last part of main result is proved in Section X. Immediate extensions of the main result to some other interesting system models are presented in the concluding Section XI.

## II. SYSTEM MODEL

We consider the problem of dynamically allocating a time-varying server to two queues. Each Queue $i \in I = \{1, 2\}$ is fed by an independent arrival process $(\boldsymbol{A}_i(t), \ t = 0, 1, \cdots)$ that is i.i.d. over $t$ and where $\boldsymbol{A}_i(t) \in \mathbb{Z}_+$ denotes the number of packets arriving in time slot $[t, t+1)$. We assume that the arrivals are bounded, i.e., $\boldsymbol{A}_i(\cdot) \leq C$ for some finite $C > 0$. Let $\boldsymbol{A}(t) = (\boldsymbol{A}_i(t), \ i \in I)$, and vector $\overline{\lambda} = E[\boldsymbol{A}(0)]$ denote the mean arrivals to the queues. We use bold face, e.g., $(\boldsymbol{A}(t), \ t = 0, 1, \cdots)$, to denote the random process and plain font, e.g., $(A(t), \ t = 0, 1, \ldots)$, to denote a realization of the process.

---

[2]By the "desired" mode of max-queue overflow we mean the mode which gives the lower bound on the asymptotic probability of max-queue overflow under *any* scheduler, as given in [15]; the likely mode of max-queue overflow under the Exp rule is indeed the desired one.

A server with randomly varying service rates is available to the two queues and modeled as follows. The server has a time-varying state that is modeled by an i.i.d. random process $(\boldsymbol{m}(t),\ t = 0, 1, \cdots)$, where $\boldsymbol{m}(t) \in \mathcal{M} = \{1, 2, \cdots, M\}$ for some finite $M > 0$ denotes the state of the server over $[t, t+1)$, and is drawn from distribution $\pi = (\pi_1, \cdots, \pi_M) > 0$. Associated with each server state $m \in \mathcal{M}$ is a vector $\mu^m \in \mathbb{Z}_+^2$. When in state $m$ over a time slot, the server can either serve at most $\mu_1^m$ packets from Queue 1, or at most $\mu_2^m$ packets from Queue 2. The scheduling problem is thus to allocate the server to one of the queues for each time slot such that a given optimality criterion is met. This will be formally described later.

At any integer time $t$, $\boldsymbol{Q}(t) = (\boldsymbol{Q}_i(t),\ i \in I) \in \mathbb{Z}_+^2$ is a random vector, where $\boldsymbol{Q}_i(t)$ denotes the number of packets in the $i^{th}$ queue at the end of time slot $[t-1, t)$. Let $S_t$ denote a system sample path up to time $t$, i.e., $S_t \equiv \big((m(\tau), Q(\tau), A(\tau-1)),\ \tau \leq t\big)$ and $\mathcal{S}_t$ denote the space of all feasible realizations $S_t$. Let $i_t^* : \mathcal{S}_t \to I$ denote the queue scheduled to receive service during time slot $[t, t+1)$, i.e., we assume that the system sample path $S_t$ is available for making the scheduling decision for time slot $[t, t+1)$. The evolution of the queue process under the scheduling decision $i_t^*(S_t)$ is given by,

$$\boldsymbol{Q}_i(t+1) = \big(\boldsymbol{Q}_i(t) - \mu_i^{\boldsymbol{m}(t)} \mathbb{1}_{\{i_t^*(S_t)=i\}}\big)^+ + \boldsymbol{A}_i(t)\ .$$

The sequence of functions $(i_t^*(\cdot),\ t = 0, 1, \ldots)$ is called a *scheduler* or *scheduling policy*. It is easy to see that under a static state-feedback scheduler, i.e., one where $i_t^*(S_t) = i^*\big(Q(t), m(t)\big)$, the process $(\boldsymbol{Q}(t),\ t = 0, 1, \ldots)$ forms a discrete time Markov chain on $\mathbb{Z}_+^2$.

In the sequel, we will extend the domain of all discrete time processes and functions to continuous time: a function originally defined on integer times has the same value at any real $t$ that it takes at $\lfloor t \rfloor$. All such processes and functions lie in the space of *real-valued right continuous functions with left limits*, denoted by $\mathcal{D}$. We assume that $\mathcal{D}$ is endowed with the topology of uniform convergence over compact sets (u.o.c), and the $k$-times product space $\mathcal{D}^k$ with the product topology. Lastly, let $(\Omega, \mathcal{F}, P)$ be a probability space that is large enough to define all the random processes in this paper.

For a given weight vector $b = (b_1, b_2) > 0$, we are interested in finding a scheduler which, informally speaking, minimizes the tail of the distribution of weighted sum-queue $\sum_{i \in I} b_i \boldsymbol{Q}_i(\cdot)$. We will show that under a static state-feedback scheduler, namely the p-Log rule described in Section IV, the asymptotic probability of weighted sum-queue overflow in the steady state, i.e.,

$$\limsup_{n \to \infty} \frac{1}{n} \log P\left(\sum_{i \in I} b_i \boldsymbol{Q}_i(0) \geq n\right)\ ,$$

is *minimized* (see Theorem 1 for a formal statement.) The p-Log rule depends only on weight vector $b$ and does not require any knowledge of arrival or server-state distributions.

*Capacity region*

For each server state $m \in \mathcal{M}$, let $V^m \in \mathbb{R}_+^2$ denote the convex hull of vertices $(0, 0)$, $(0, \mu_2^m)$, and $(\mu_1^m, 0)$. Then, conditional on the server being in state $m$, the average service jointly offered to the two queues under any scheduling rule (such that the average exists) lies in the *triangle* $V^m$. Define the capacity region $V_\pi$ as the set of average service vectors offered to the two queues under all possible scheduling rules, then $V_\pi$ is a convex polyhedron given by the weighted Minkowski sum of regions $V^m$, i.e.,

$$\begin{aligned} V_\pi &= \pi_1 V^1 \oplus \cdots \oplus \pi_M V^M\ , \\ &= \left\{\sum_{m \in \mathcal{M}} \pi_m v(m)\ :\ v(m) \in V^m,\ m \in \mathcal{M}\right\}\ . \end{aligned} \quad (1)$$

See Fig. 1-$a$ for a graphical illustration of capacity region: the server has $M = 6$ states with some distribution $\pi$ and vectors $\mu^m$ in set $\{(1, 4), (3, 4), (1, 1), (4, 3), (4, 1), (1, 0)\}$.

Let $\{r_1, \cdots, r_{M'}\}$ for some $M' \leq M$ be the set of strictly positive and finite slopes of the outer normal vectors to the facets of capacity region $V_\pi$. The slopes are indexed such that $0 < r_1 < r_2 < \ldots < r_{M'} < \infty$. Also, let $r_0 = 0$ and $r_{M'+1} = \infty$. For example, see Fig. 1-$a$ for a depiction of a capacity region with $M' = 5$ facets with outer normal slopes in $(0, \infty)$. Finally, let $\hat{V}_\pi = \{\hat{v}^1, \cdots, \hat{v}^{M'+1}\}$ denote the set of maximal vertices of the capacity region $V_\pi$. The vertices in set $\hat{V}_\pi$ are indexed such that $\hat{v}_1^1 > \hat{v}_1^2 > \ldots > \hat{v}_1^{M'+1}$, i.e., the vertex $\hat{v}^m$ lies at the intersection of the facets with outer normal slopes $r_{m-1}$ and $r_m$; e.g., see vertex $\hat{v}^2$ in Fig. 1-$a$.

We assume that there exists a $v \in V_\pi$ such that $\overline{\lambda} < v$, which is a sufficient condition for stabilizability of the queues [3]. If the above condition is met, then there exists at least one static state feedback scheduler under which the Markov chain $(\boldsymbol{Q}(t),\ t = 0, 1, \ldots)$ is ergodic.

*Remark 1:* Since we assumed that the server can be allocated to at most one queue per time slot, the region $V^m$ is obtained by taking the convex hull of service vectors in the set $\{(0, 0), (\mu_1^m, 0), (0, \mu_2^m)\}$. However, we can relax this assumption and allow the server to be shared between the two queues during a time slot. That is, we can associate with each server state $m$ a set of $k_m$ service vectors $\{(\mu_1^m(1), \mu_2^m(1)), \cdots, (\mu_1^m(k_m), \mu_2^m(k_m))\}$, and allow the server to operate at any one of these service vectors. In this more general case, each region $V^m$ will be an arbitrary convex polyhedron obtained by taking the convex hull of all feasible service vectors associated with server state $m$. For example, $V^m$ can be information theoretic polymatroids as in [9]. The optimality results presented in this paper will still hold with such a relaxation; see Section XI for some details.

### III. THROUGHPUT-OPTIMAL SCHEDULERS AND RADIAL SUM-RATE MONOTONICITY

The throughput-optimal schedulers MaxWeight, Exp rule, and Log rule mentioned in the introduction are all static state-feedback. We formally define them next. Let the vector fields $h^{mw}(\cdot), h^{exp}(\cdot)$ and $h^{log}(\cdot)$ on $\mathbb{R}_+^2$ be given as follows:

for all $x \in \mathbb{R}_+^2$,

$$h^{mw}(x) = \left(b_i x_i^\alpha, \ i \in I\right),$$
$$h^{exp}(x) = \left(b_i \exp\left(\frac{a_i x_i}{c + (0.5(a_1 x_1 + a_2 x_2))^\eta}\right), \ i \in I\right),$$
$$h^{log}(x) = \left(b_i \log\left(1 + a_i x_i\right), \ i \in I\right),$$

for any fixed positive $b_i$'s, $a_i$'s, $\alpha$, $c$, and $0 < \eta < 1$. Then, when the system is in state $(\boldsymbol{Q}(t), \boldsymbol{m}(t)) = (Q, m) \in \mathbb{Z}_+^2 \times \mathcal{M}$, the MaxWeight scheduler serves a queue $i^*_{mw}$ given by,

$$i^*_{mw}(Q, m) \in \arg\max_{i \in I} h_i^{mw}(Q) \mu_i^m, \quad (2)$$

augmented with any fixed tie-breaking rule. The Exp rule $i^*_{exp}$ and the Log rule $i^*_{log}$ are defined similarly by substituting $h^{exp}$ and $h^{log}$ respectively in place of $h^{mw}$. Indeed numerous vector field based throughput-optimal schedulers can be engineered so as to respond differently to the disparity among the users' queue lengths, i.e., make different tradeoffs between *service rate maximization* and *queue balancing*. For reference see, e.g., [7], which gives necessary and sufficient conditions for a vector field based scheduler to be throughput-optimal.

We refer to a scheduler as *radial sum-rate monotone* if, as the queues scale up linearly, the scheduling rule allocates the server in a manner that de-emphasizes queue-balancing in favor of greedily maximizing the current service rate. More formally, let $v(Q) \in V_\pi$ be the vector of average service offered to the queues under a static state-feedback scheduler $i^*$, conditional on queue state being $Q$, i.e.,

$$v(Q) = \left(E[\mu_i^{\boldsymbol{m}} \mathbb{1}_{\{i^*(Q,\boldsymbol{m})=i\}}], \ i \in I\right), \quad (3)$$

where expectation is with respect to (random) $\boldsymbol{m}$ drawn from distribution $\pi$.

*Definition 1:* A scheduling policy $i^*$ is *radial sum-rate monotone* with respect to weight vector $b > 0$ if for any $Q$ and scalar $\theta$ such that $\theta Q \in \mathbb{Z}_+^n$, the weighted sum of expected offered service, $\langle b, v(\theta Q) \rangle$, is an increasing function of $\theta$, and

$$\lim_{\theta \to \infty} \langle b, v(\theta Q) \rangle = \max_y \left(\langle b, y \rangle \mid y \in V_\pi \text{ and } y_i = 0 \text{ if } Q_i = 0\right).$$

Let $v^{mw}(Q)$ denote the expected service vector under the MaxWeight scheduler, i.e., the vector obtained by substituting $i^*_{mw}$ for $i^*$ in (3); similarly, let $v^{exp}(Q)$ and $v^{log}(Q)$ denote the expected service vectors under the Exp and Log rule respectively. Also, fix a weight vector $b > 0$. Next, under schedulers Maxweight, Exp rule, and Log rule respectively, we will identify the sets $\mathcal{S}_0^{mw}, \mathcal{S}_0^{exp}$, and $\mathcal{S}_0^{log}$ given *approximately*[3] by,

$$\mathcal{S}_0^{(\cdot)} \approx \left\{Q \in \mathbb{Z}_+^2 : v^{(\cdot)}(Q) \in \arg\max_{y \in V_\pi} \langle y, b \rangle\right\},$$

in words, the set of queue states such that the expected service vector $v^{(\cdot)}(Q)$ has the maximum weighted sum with respect to weight vector $b$. For example, for the capacity region shown in Fig. 1-$a$ and weight vector $b = (1,1)$, Fig. 1-$b$–$d$ illustrates the sets $\mathcal{S}_0^{mw}, \mathcal{S}_0^{exp}$, and $\mathcal{S}_0^{log}$. The set $\mathcal{S}_0^{mw}$ is a cone, the set $\mathcal{S}_0^{exp}$ a cylinder with gradually increasing diameter, and the set $\mathcal{S}_0^{log}$ resembles a French horn. The set $\mathcal{S}_0^{log}$ is such that for any $Q > 0$, we have $\theta Q \in \mathcal{S}_0^{log}$ for all $\theta$ large enough, indicating that the Log rule is radial sum-rate monotone. The MaxWeight and Exp rule are not radial sum-rate monotone. In fact, Exp rule is the 'opposite' of radial sum-rate monotone in that, for any $Q$ such that $a_1 Q_1 \neq a_2 Q_2$, we have $\theta Q \notin \mathcal{S}_0^{exp}$ for all $\theta$ large enough. A formal description of the sets $\mathcal{S}_0^{(\cdot)}$ is as follows.

---

[3]Under the formal definition of set $\mathcal{S}_0^{(\cdot)}$ given later, the queue states on the boundary of set $\{Q \in \mathbb{Z}_+^2 : v^{(\cdot)}(Q) \in \arg\max_{y \in V_\pi} \langle y, b \rangle\}$ may or may not lie in $\mathcal{S}_0^{(\cdot)}$.

The vector $h^{(\cdot)}(Q)$ can be shown to be an outer normal vector to the capacity region $V_\pi$ at point $v^{(\cdot)}(Q)$ for each scheduler $(\cdot) \in \{mw, exp, log\}$, i.e.,

$$v^{(\cdot)}(Q) \in \arg\max_{y \in V_\pi} \left\langle y, h^{(\cdot)}(Q) \right\rangle, \quad (4)$$

See, e.g., Lemma 2.1 of [16]. Recall the set of outer normal slopes $\{r_0, r_1, \cdots, r_{M'}, r_{M'+1}\}$, and let $r_k$ be the largest slope strictly less than $b_2/b_1$ (i.e. the slope of vector $b$) and $r_l$ the smallest slope strictly greater than $b_2/b_1$; thus if $r_{k+1} \neq b_2/b_1$ then $l = k+1$, otherwise $l = k+2$. For example, in Fig. 1-$a$, for weight vector $b = (1,1)$, we have $k = 2$ and $l = 4$. For each scheduler $(\cdot) \in \{mw, exp, log\}$ we define the set,

$$\mathcal{S}_0^{(\cdot)} = \left\{x \in \mathbb{R}_+^2 : r_k < \frac{h_2^{(\cdot)}(x)}{h_1^{(\cdot)}(x)} < r_l\right\}.$$

Then, by (4), for any $Q \in \mathcal{S}_0^{(\cdot)} \cap \mathbb{Z}_+^2$, we must have $v^{(\cdot)}(Q) \in \arg\max_{y \in V_\pi} \langle y, b \rangle$. Each region $\mathcal{S}_0^{(\cdot)}$ is bounded by *switching curves* given by,

$$\left\{x \in \mathbb{R}_+^2 : \frac{h_2^{(\cdot)}(x)}{h_1^{(\cdot)}(x)} = r_k\right\} \text{ and } \left\{x \in \mathbb{R}_+^2 : \frac{h_2^{(\cdot)}(x)}{h_1^{(\cdot)}(x)} = r_l\right\}.$$

For any queue state $Q$ lying on these switching curves, the argmax in (4) is not unique and whether $v^{(\cdot)}(Q)$ lies in the set $\arg\max_{y \in V_\pi} \langle y, b \rangle$ depends on the tie breaking rule associated with (2). For each $m \in \{1, \cdots, k, l+1, \cdots, M'+1\}$, we can also define a set,

$$\mathcal{S}_m^{(\cdot)} = \left\{x \in \mathbb{R}_+^2 : r_{m-1} < \frac{h_2^{(\cdot)}(x)}{h_1^{(\cdot)}(x)} < r_m\right\}.$$

Then, again by (4), for any $Q \in \mathcal{S}_m^{(\cdot)} \cap \mathbb{Z}_+^2$, we must have $v^{(\cdot)}(Q) = \hat{v}^m$. That is, each set $\mathcal{S}_m^{(\cdot)}$ for $m \in \{1, \cdots, k, l+1, \cdots, M'+1\}$ is associated with a unique vertex. The switching curve between any two regions $\mathcal{S}_m^{(\cdot)}$ and $\mathcal{S}_{m+1}^{(\cdot)}$ for $m \in \{1, \cdots, k-1, l+1, \cdots, M'\}$ is given by,

$$\left\{x \in \mathbb{R}_+^2 : \frac{h_2^{(\cdot)}(x)}{h_1^{(\cdot)}(x)} = r_m\right\},$$

and as before, for any queue state $Q$ lying on this switching curve, the argmax in (4) is not unique and whether it lies in set $\mathcal{S}_m^{(\cdot)}$ or $\mathcal{S}_{m+1}^{(\cdot)}$ depends on the tie breaking rule associated with (2). Fig. 1-$b$–$d$ also show the switching curves and sets $\mathcal{S}_m^{(\cdot)}$ under MaxWeight, Exp rule, and Log rule. In fact, for MaxWeight and Log rule, the switching curves can be given

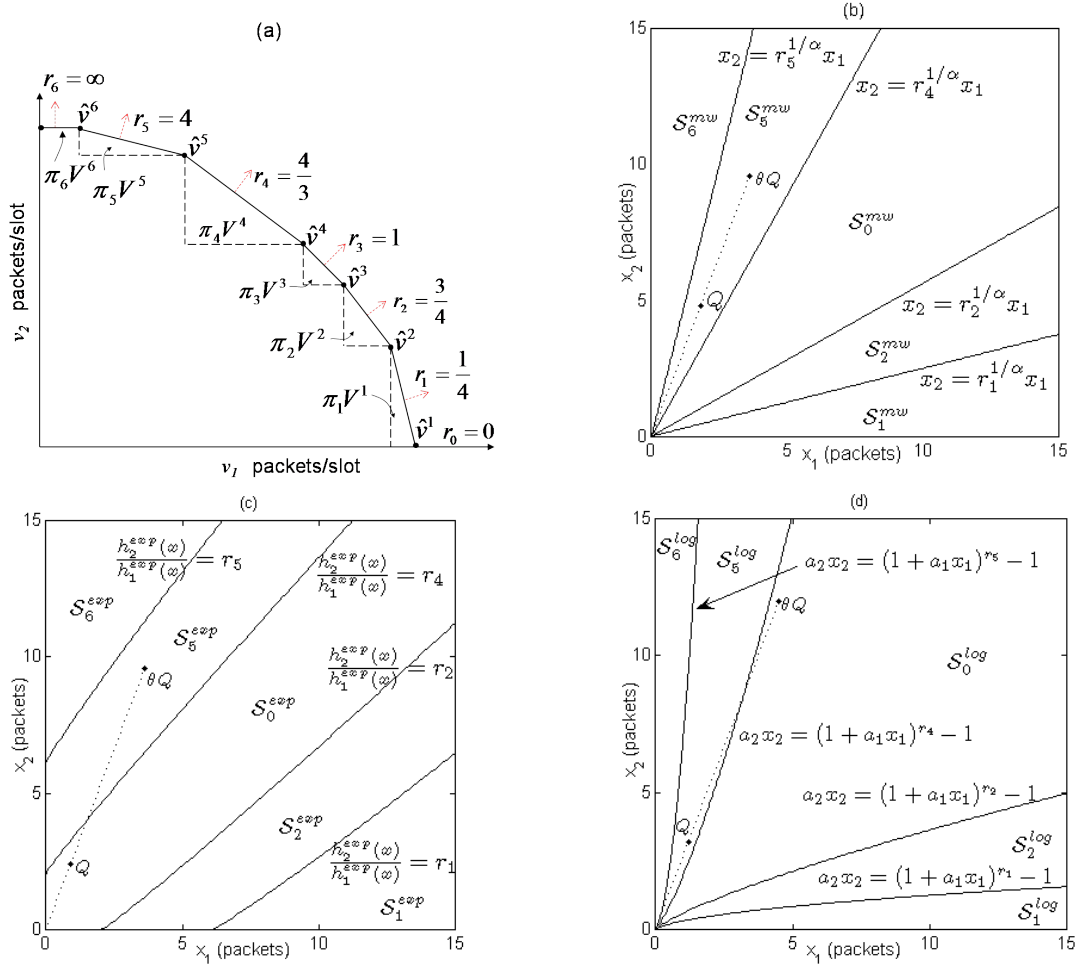

Fig. 1. (a) The capacity region for $\mu^m \in \{(1,4),(3,4),(1,1),(4,3),(4,1),(1,0)\}$, depicting Minkowski addition, outer-normal vectors and maximal vertices; the resulting switching curves under (b) MaxWeight, (c) Exp rule, (d) Log rule, for weight vector $b = (1,1)$.

in closed form, e.g., for Log rule, the switching curves are given by,

$$a_2 x_2 = (1 + a_1 x_1)^{\frac{b_1}{b_2} r} - 1 \;,$$

for $r \in \{r_1, \cdots, r_{M'}\}$.

## IV. THE PSEUDO-LOG SCHEDULING RULE

In this section we introduce a static state-feedback and radial sum-rate monotone scheduler, denoted the pseudo-Log (p-Log) rule. We subsequently show in Theorem 1 that the p-Log rule minimizes the asymptotic probability of weighted sum-queue overflow for any weight vector $b$. The p-Log rule takes weight vector $b$ as a parameter but does not require any knowledge of arrival or server-state distributions.

We will define the p-Log scheduling rule through a vector field $h = (h_1, h_2)$ on $\mathbb{R}_+^2$. For any $x \in [0,1)^2$, let $h(x) = 0$. For any $x \in \mathbb{R}_+^2 \setminus [0,1)^2$,

$$\text{if } x_1 \geq x_2, \text{ then let } \begin{cases} h_1(x) = b_1 \sqrt{x_1}, \\ h_2(x) = b_2 \min(x_2, \sqrt{x_1}); \end{cases}$$
$$\text{if } x_1 < x_2, \text{ then let } \begin{cases} h_1(x) = b_1 \min(x_1, \sqrt{x_2}), \\ h_2(x) = b_2 \sqrt{x_2}. \end{cases} \quad (5)$$

The p-Log rule is given as follows: when the queues are in state $Q \in \mathbb{Z}_+^2$ and the server in state $m \in \mathcal{M}$, then the server is allocated to queue $i^*_{pLog}(Q, m)$ given by,

$$i^*_{pLog}(Q, m) \in \arg\max_{i \in I} h_i(Q) \mu_i^m \;, \quad (6)$$

where, in the case of a tie, if $Q_1 \geq Q_2$ the server is allocated to Queue 1, otherwise to Queue 2.

Note that it is only the slope, $\frac{h_2(Q)}{h_1(Q)}$, of the vector $h(Q)$ that determines the scheduling decision, and so, for example, when $Q_1 \geq Q_2$, $Q_1 \neq 0$, the slope is given by,

$$\frac{h_2(Q)}{h_1(Q)} = \frac{b_2}{b_1} \min\left(1, \frac{Q_2}{\sqrt{Q_1}}\right) \;.$$

*Switching curves under the p-Log rule*

Let $v^{pLog}(Q)$ denote the vector of average service offered to the queues under p-Log rule, conditional on queue state being $Q$, i.e., the vector obtain by substituting $i^*_{pLog}$ for $i^*$ in (3). Recall the set of outer normal slopes $\{r_0, r_1, \cdots, r_{M'}, r_{M'+1}\}$ and, in particular, the slopes $r_k$ and $r_l$ from this set as defined in Section III. Similar to the sets $\mathcal{S}_0^{(\cdot)}$ for $(\cdot) \in \{mw, exp, log\}$,

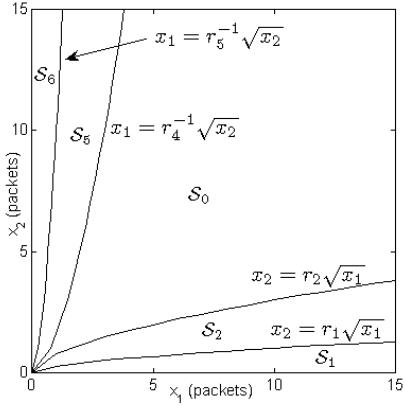

Fig. 2. The switching curves under the p-Log rule for the capacity region depicted in Fig. 1-a and weight vector $b = (1, 1)$.

we define the set $\mathcal{S}_0$ for p-Log rule as,

$$\begin{aligned}\mathcal{S}_0 &= \left\{ x \in \mathbb{R}_+^2 : r_k < \frac{h_2(x)}{h_1(x)} < r_l \right\}, \\ &= \left\{ x \in \mathbb{R}_+^2 : \frac{b_1}{b_2} r_k \sqrt{x_1} < x_2 < \left(\frac{b_1}{b_2} r_l \ x_1\right)^2 \right\}.\end{aligned}$$

Similar to the region $\mathcal{S}_0^{log}$, the region $\mathcal{S}_0$ too is shaped like a French horn, see Fig. 2 for an illustration. For any $Q \in \mathcal{S}_0 \cap \mathbb{Z}_+^2$, we have $v^{pLog}(Q) \in \arg\max_{y \in V_\pi} \langle y, b \rangle$. Therefore, we will refer to the region $\mathcal{S}_0$ as the weighted-max-sum rate region (with respect to weight vector $b$). Moreover, for any $Q > 0$, we have $\theta Q \in \mathcal{S}_0$ for all $\theta$ large enough, thus indicating that p-Log rule is radial sum-rate monotone. Also, for each $m \in \{1, \cdots, k, l+1, \cdots, M'+1\}$, we define a set $\mathcal{S}_m$ as,

$$\mathcal{S}_m = \left\{ x \in \mathbb{R}_+^2 : r_{m-1} < \frac{h_2(x)}{h_1(x)} < r_m \right\}.$$

Then for any $Q \in \mathcal{S}_m \cap \mathbb{Z}_+^2$, we have $v^{pLog}(Q) = \hat{v}^m$. All switching curves in the half plane $\{x_1 \geq x_2\}$ are given by,

$$x_2 = \frac{b_1}{b_2} r \sqrt{x_1},$$

for $r \in \{r_0, \cdots, r_k\}$ and $x_1 \geq 1$. Similarly, all switching curves in the half plane $x_2 > x_1$ are given by,

$$x_1 = \frac{b_2}{b_1} r^{-1} \sqrt{x_2},$$

for $r \in \{r_l, \cdots, r_{M'+1}\}$ and $x_2 \geq 1$. We will refer to the collection of switching curves and regions $\mathcal{S}_m$ as a *partition* of $\mathbb{R}_+^2$ (or the queue state space $\mathbb{Z}_+^2$) under the p-Log rule. It will be useful to note that this partition, as well as the partitions under MaxWeight, Exp rule, and Log rule, depend only on the set of vectors $\{\mu^1, \cdots, \mu^M\}$ associated with the $M$ server states and not on the distribution $\pi = (\pi_1, \cdots, \pi_M) > 0$ over these states.

## V. MAIN RESULT

The following three-part theorem summarizes the main results of this paper. It includes a lower bound on the tail of the weighted sum-queue overflow probability, an upper bound on the same, and the optimality of the p-Log rule. The first part is proved in Section VII, the second in Section VIII (and Appendix), while the last in Sections IX and X.

*Theorem 1:* For the system model detailed in Section II, the following hold.
(i) Given a weight vector $b = (b_i : b_i > 0, i \in I)$, there exists finite $T_0 > 0$ such that for any $t > T_0$ and under *any scheduling rule* starting in any initial state $Q(0)$, we have the following lower bound,

$$\liminf_{n \to \infty} \frac{1}{n} \log P\left( \sum_{i \in I} b_i Q_i(nt) \geq n \right) \geq -J_* ,$$

where $J_*$ is defined in Section VII.
(ii) For a stabilizable system, under p-Log scheduling rule the process $(Q(t), \ t = 0, 1, \ldots)$ forms an ergodic Markov chain, and we have the following upper bound for a random vector $Q$ drawn from the stationary distribution of the Markov chain,

$$\limsup_{n \to \infty} \frac{1}{n} \log P\bigl( \sum_{i \in I} b_i Q_i \geq n \bigr) \leq -J_{**} ,$$

where $J_{**}$ is defined in Section VIII.
(iii) The p-Log rule maximizes the asymptotic exponential decay rate of the weighted sum-queue distribution, i.e.,

$$J_{**} = J_* .$$

*Remark 2:* Since a scheduler can, in principle, be non-stationary, the lower bound in *(i)* is expressed in a more general form than the upper bound in *(ii)* which is specific to a static state-feedback scheduler, namely the p-Log rule. For stationary schedulers under which the process $(Q(t), \ t = 0, 1, \ldots)$ forms an ergodic Markov chain, the lower bound in *(i)* also implies the same lower bound under the steady state distribution of the Markov chain. This is because the lower bound will hold if the initial state $Q(0)$ were random and drawn from the steady state distribution.

## VI. FLUID-SCALED PROCESSES AND LARGE DEVIATION PRINCIPLES

In this section, we define sequences of fluid-scaled processes and functions, and a Large Deviation Principle [17] on those sequences, which is used in proving Theorem 1. Define the cumulative arrivals process $\boldsymbol{F} = \bigl(\boldsymbol{F}(t) = (\boldsymbol{F}_i(t), \ i \in I), t \geq 0\bigr)$ obtained from the process $(\boldsymbol{A}(t), \ t \geq 0)$ as,

$$\boldsymbol{F}_i(t) = \sum_{k=0}^{\lfloor t-1 \rfloor} \boldsymbol{A}_i(k) ,$$

and cumulative time the channel is in each state $\boldsymbol{G} = \bigl(\boldsymbol{G}(t) = (\boldsymbol{G}_m(t), \ m \in \mathcal{M}), \ t \geq 0 \bigr)$ obtained from the process $(\boldsymbol{m}(t), \ t \geq 0)$ as,

$$\boldsymbol{G}_m(t) = \sum_{k=0}^{\lfloor t-1 \rfloor} \mathbb{1}_{\{\boldsymbol{m}(k) = m\}} .$$

The triplet $(Q, F, G)$, where $Q = (Q(t), \ t \geq 0)$ is the queue sample path (under a fixed scheduling rule) corresponding to the sample paths $(F, G)$ and initial state $Q(0)$, denotes a realization of the system $(Q, F, G)$. For each $n = 0, 1, \ldots$, let $(Q^{(n)}, F^{(n)}, G^{(n)})$ denote an independent and identically distributed system. We define a corresponding sequence of fluid-scaled processes, denoted by $(q^{(n)}, f^{(n)}, g^{(n)})$, as,

$$q^{(n)} = \left(q^{(n)}(t),\ t \geq 0\right) = \left(\frac{1}{n}Q^{(n)}(nt),\ t \geq 0\right),$$

with $f^{(n)}$ and $g^{(n)}$ similarly defined.

The arrival and service processes are i.i.d. and bounded and, therefore, satisfy large deviation principles [17]. In particular, for each $i \in I$, define for any scalar $\lambda_i \geq 0$ the rate function $L_i(\cdot)$ for the sequence $f_i^{(n)}(1)$ as,

$$L_i(\lambda_i) = \sup_{\theta \geq 0} \left(\theta \lambda_i - \log E\left[e^{\theta A_i(1)}\right]\right),$$

where $L_i(\cdot) = \infty$ over $(-\infty, 0)$ and $(C, \infty)$. Also, for any vector $\lambda \in \mathbb{R}_+^2$, let,

$$L_{(f)}(\lambda) = \sum_{i \in I} L_i(\lambda_i).$$

For any probability distribution $\gamma = (\gamma_m,\ m \in \mathcal{M})$, we define the relative entropy $L_{(g)}(\cdot)$ of $\gamma$ with respect to distribution $\pi$ as,

$$L_{(g)}(\gamma) = \sum_{m \in \mathcal{M}} \gamma_m \log \frac{\gamma_m}{\pi_m}.$$

where $L_{(g)}(\cdot) = \infty$ everywhere outside the standard simplex in $\mathbb{R}_+^M$. Now consider any functions $(f, g) \in \mathcal{D}^{2+M}$. If $(f, g)$ are absolutely continuous, then they are differentiable *a.e.* and we let $(f'(t), g'(t)) = \frac{d}{dt}(f(t), g(t))$. For any $t > 0$, if $(f(0), g(0)) = 0$ and $(f, g)$ are absolutely continuous on the interval $[0, t]$, then let,

$$J_t(f, g) = \int_0^t L_{(f)}\bigl(f'(s)\bigr) + L_{(g)}\bigl(g'(s)\bigr)\ ds,$$

otherwise $J_t(f, g) = \infty$. The functional $J_t(f, g)$ is referred to as the cost of the trajectories $(f, g)$ over the time interval $[0, t]$. The following is a form of Borovkov/Mogulskii's theorem [17].

*Proposition 1:* For any fixed $T > 0$, consider a sequence in $n$ of the fluid-scaled processes $(f^{(n)}, g^{(n)}) = ((f^{(n)}(t), g^{(n)}(t)),\ t \in [0, T])$, then for any measurable $B \subseteq \mathcal{D}^{2+M}$, we have that,

$$-\inf_{(f,g)} \left\{J_T(f, g) | (f, g) \in B^\circ\right\}$$
$$\leq \liminf_{n \to \infty} \frac{1}{n} \log P((f^{(n)}, g^{(n)}) \in B)$$
$$\leq \limsup_{n \to \infty} \frac{1}{n} \log P((f^{(n)}, g^{(n)}) \in B)$$
$$\leq -\inf_{(f,g)} \left\{J_T(f, g) | (f, g) \in \overline{B}\right\},$$

where, $B^\circ$ and $\overline{B}$ denote the interior and closure of set $B$ respectively.

Let $u(n) = \lceil n^\alpha \rceil$ for some fixed $\alpha \in (0, 0.5)$. For any function $d \in \mathcal{D}^{2+M}$, let $U^n d$ denote the piece-wise linear function obtained by linear interpolation over samples $\left(d(\frac{ku(n)}{n}),\ k = 0, 1, \ldots\right)$. The following upper bound is Stolyar's refinement of Mogulskii's theorem and was first introduced in [15].

*Proposition 2:* For any fixed $T > 0$, consider a sequence in $n$ of the fluid-scaled processes $(f^{(n)}, g^{(n)}) = ((f^{(n)}(t), g^{(n)}(t)),\ t \in [0, T])$. Suppose, for each $n$ there is a fixed measurable $B^{(n)} \subseteq \mathcal{D}^{2+M}$, that is a subset of the set of feasible realizations of $(f^{(n)}, g^{(n)})$ in $[0, T]$. Then,

$$\limsup_{n \to \infty} \frac{1}{n} \log P\left((f^{(n)}, g^{(n)}) \in B^{(n)}\right) \qquad (7)$$
$$\leq -\liminf_{n \to \infty} \inf_{(f,g)} \left\{J_{T^{(n)}} U^n(f, g) | (f, g) \in B^{(n)}\right\},$$

where $T^{(n)} = \frac{u(n)}{n} \lfloor \frac{nT}{u(n)} \rfloor$.

Note that by contrast with Proposition 1, $\{B^{(n)}\}$ in Proposition 2 corresponds to a sequence of sets of scaled feasible trajectories, and the bound on the right side is an infimum over the cost of sampled and linearly interpolated scaled trajectories. Thus this proposition provides a refinement allowing, for example, the consideration of large deviations for sets $\{B^{(n)}\}$ which can distinguish among the trajectories in the set $\{f^{(k)} \in \mathcal{D}^2 : f^{(k)}$ converges u.o.c. to a Lipschitz $f\}$. That is to say, even if all the fluid scaled trajectories in a set converge to the same limiting trajectory, the events $B^{(n)}$ can be defined to include only a subset of these trajectories.

Also introduced in [15] is a notion of generalized fluid sample path (GFSP) which naturally appears when applying the bound in Proposition 2; we describe this next. Consider a sequence of realizations $(q^{(n)}, f^{(n)}, g^{(n)})\ n = 1, 2, \ldots$ such that along some subsequence (still denoted by $\{n\}$), we have u.o.c convergence

$$(q^{(n)}, f^{(n)}, g^{(n)}) \to (q, f, g)$$

to some Lipschitz continuous functions $(q, f, g)$, and u.o.c convergence

$$\overline{J}^{(n)} = \left(\overline{J}_t^{(n)},\ t \geq 0\right),$$
$$= \left(J_{t^{(n)}} U^n(f^{(n)}, g^{(n)}),\ t \geq 0\right) \to \overline{J} = \left(\overline{J}_t,\ t \geq 0\right)$$

to some non-negative increasing Lipschitz continuous function $\overline{J}$. Then the following construct is called a GFSP,

$$\psi = \left[\left(q^{(n)}, f^{(n)}, g^{(n)}\right),\ \overline{J}^{(n)},\ n = 0, 1, \cdots;\ (q, f, g);\ \overline{J}\right]$$

and the function $\overline{J}$ is referred to as the refined cost function of the GFSP. Since this construct contains not only the limiting trajectory $(q, f, g)$ but also the sequence converging to it, the construct is useful when one needs to *zoom in* and study the limiting trajectories $(q, f, g)$ at a finer-than-fluid scaling. Moreover, we will see that for the events $B^{(n)}$ of interest in this paper, the bound in Proposition 2 reduces to an infimum over the refined cost of a well-defined set of GFSPs.

## VII. Lower bound on overflow probability under any scheduling rule

For any distribution $\gamma$ on the set $\mathcal{M}$ of server states, let $V_\gamma$ denote the corresponding capacity region, see (1). Let $b = (b_1, b_2) > 0$ be the given weight vector. For any vector $\lambda \in \mathbb{R}_+^2$, let $v^*(\lambda, \gamma)$ denote a service vector such that,

$$v^*(\lambda, \gamma) \in \arg\max_v \{\langle b, v \rangle : v \leq \lambda, \; v \in V_\gamma\} , \qquad (8)$$

where, if the argmax is not unique, then $v^*(\lambda, \gamma)$ can be set to any maximizer. For example, see Fig. 3 depicting $v^*(\lambda, \pi)$ for weight vector $b = (1, 1)$ and two hypothetical vectors $\lambda^{(1)}$ and $\lambda^{(2)}$ lying outside capacity region $V_\gamma$. The interpretation is as follows: if the arrival process were to exhibit an empirical mean of $\lambda$ and the server state were to exhibit an empirical distribution $\gamma$, then serving the queues according to the service vector $v^*(\lambda, \gamma)$ minimizes the rate of weighted-sum-queue build-up, i.e., $\langle b, \lambda - v^*(\lambda, \gamma)\rangle$.

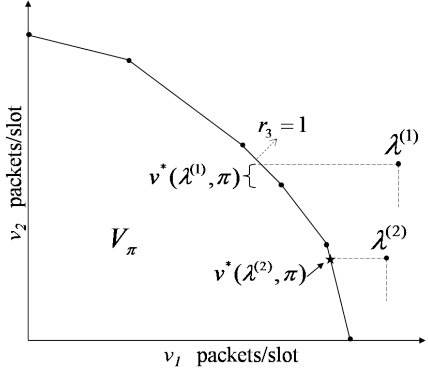

Fig. 3. For weights $b_i = 1$, the *optimal* capacity vector $v^*(\lambda^{(2)}, \pi)$ is unique, whereas, vector $v^*(\lambda^{(1)}, \pi)$ is any point on the annotated part of max-sum-rate face of region $V_\pi$.

Finally, we define the minimum cost (per unit increase of the weighted-sum-queue) $J_*$ as,

$$J_* = \min_{\gamma, \lambda} \frac{L_{(g)}(\gamma) + L_{(f)}(\lambda)}{\langle b, \lambda - v^*(\lambda, \gamma)\rangle} , \qquad (9)$$

### A. Proof of Theorem 1-(i)

Let $(\lambda^*, \gamma^*)$ be a point that achieves the minimum in (9), and let $T_0 = \langle b, \lambda^* - v^*(\lambda^*, \gamma^*)\rangle^{-1}$. We will show that regardless of the scheduling rule, all realizations $(f, g)$ sufficiently *close* to $(\lambda^* t, \gamma^* t)$ over the interval $[0, T_0]$– and thus having cost $J_{T_0}(f, g)$ close to $J_*$– lead to an overflow at time $T_0$.

Let $||.||$ denote the $L_\infty$ norm. For any $\epsilon > 0$, define a set of trajectories over the interval $[0, T_0]$,

$$\begin{aligned} B_\epsilon = \; & \left\{ (f,g) \in \mathcal{D}^{2+M} : (f,g) \text{ is absolutely continuous;} \right. \\ & \forall t \in [0, T_0], \\ & ||L_{(f)}(f'(t)) - L_{(f)}(\lambda^*)|| \leq \frac{\epsilon}{2T_0}; \\ & ||L_{(g)}(g'(t)) - L_{(g)}(\gamma^*)|| \leq \frac{\epsilon}{2T_0}; \\ & \forall u \in V_{g'(t)}, \; \inf_{v \in V_{\gamma^*}} ||u - v|| \leq \frac{\epsilon}{5T_0}; \text{ and} \quad (10) \\ & \left. ||\lambda^* - \frac{1}{T_0}f(T_0)|| \leq \frac{\epsilon}{5T_0} \right\} \qquad (11) \end{aligned}$$

The set $B_\epsilon$ is measurable, compact, and the cost for any $(f, g) \in B_\epsilon$ satisfies $||J_{T_0}(f, g) - J_*|| \leq \epsilon$. Next, we will show that any sequence of realizations $(q^{(n)}, f^{(n)}, g^{(n)})$, such that $q^{(n)}(0) = 0$ and $(f^{(n)}, g^{(n)})$ converge uniformly in $[0, T_0]$ to some $(f, g) \in B_\epsilon$, must have $\sum_{i \in I} b_i q_i^{(n)}(T_0) > 1 - 2\epsilon$ for all sufficiently large $n$. Therefore,

$$\begin{aligned} \liminf_{n \to \infty} & \frac{1}{n} \log P\left(\sum_{i \in I} b_i \boldsymbol{q}_i^{(n)}(T_0) > 1 - 2\epsilon\right) \\ & \geq \liminf_{n \to \infty} \frac{1}{n} \log P\left((\boldsymbol{f}^{(n)}, \boldsymbol{g}^{(n)}) \in B_\epsilon\right) \geq -J_* , \end{aligned}$$

where the rightmost inequality follows from the Mogulskii's theorem (Proposition 1).

Let $\lambda^{(n)} = \frac{1}{T_0} f^{(n)}(T_0)$, and $u^{(n)}$ denote the average service vector seen by the queues; that is to say, $u_i^{(n)} T_0$ is equal to the number of packets served from the $i^{th}$ queue over the interval $[0, T_0]$. By (11) in the definition of $B_\epsilon$, for all $n$ sufficiently large,

$$\lambda^* - \left(\frac{\epsilon}{4T_0}, \frac{\epsilon}{4T_0}\right) \overset{(i)}{\leq} \lambda^{(n)} \overset{(ii)}{\leq} \lambda^* + \left(\frac{\epsilon}{4T_0}, \frac{\epsilon}{4T_0}\right) , \quad (12)$$

and by (10) there exists a $v \in V_{\gamma^*}$ such that

$$v - \left(\frac{\epsilon}{4T_0}, \frac{\epsilon}{4T_0}\right) \overset{(i)}{\leq} u^{(n)} \overset{(ii)}{\leq} v + \left(\frac{\epsilon}{4T_0}, \frac{\epsilon}{4T_0}\right) . \quad (13)$$

We must also have $u^{(n)} \leq \lambda^{(n)}$, since the total service cannot exceed the total arrivals. This, along with inequalities (12-ii) and (13-i), implies $v \leq \lambda^* + \left(\frac{\epsilon}{2T_0}, \frac{\epsilon}{2T_0}\right)$. Without loss of generality, we assume that the weight vector $b$ is normalized to have $\sum_{i \in I} b_i = 2$. Then using (13-ii), we get,

$$\begin{aligned} \left\langle b, u^{(n)}\right\rangle & \leq \langle b, v\rangle + \frac{\epsilon}{2T_0} , \\ & \leq \max\left(\langle b, y\rangle : y \in V_{\gamma^*}, y \leq \lambda^* + \left(\frac{\epsilon}{2T_0}, \frac{\epsilon}{2T_0}\right)\right) \\ & \quad + \frac{\epsilon}{2T_0} , \\ & \leq \langle b, v^*(\lambda^*, \gamma^*)\rangle + \frac{\epsilon}{T_0} . \end{aligned} \quad (14)$$

Finally, using (12-i) and (14), we have that,

$$\begin{aligned} \left\langle b, q^{(n)}(T_0)\right\rangle & = \left(\left\langle b, \lambda^{(n)}\right\rangle - \left\langle b, u^{(n)}\right\rangle\right) T_0 , \\ & > \left(\langle b, \lambda^*\rangle - \langle b, v^*(\lambda^*, \gamma^*)\rangle\right) T_0 - 2\epsilon , \\ & = 1 - 2\epsilon . \end{aligned}$$

In order to obtain an overflow at any time after $T_0$ while still incurring a cost close to $J_*$, the trajectory $(\lambda^* t, \gamma^* t)$ can be prepended with the zero-cost trajectory for an appropriate amount of time. ∎

## VIII. UPPER BOUND ON OVERFLOW PROBABILITY UNDER THE p-LOG RULE

Recall the notion of a GFSP and its refined cost function from Section VI. Let $J_{**}$ denote the lowest refined cost of a GFSP that, under p-Log rule, raises $\sum_{i \in I} b_i q_i(t)$ to 1 from the initial state $q(0) = 0$, i.e.,

$$J_{**} = \inf_{t \geq 0} J_{**, t} , \qquad (15)$$

where,
$$J_{**,t} = \inf_\psi \left\{ \overline{J}_t | \psi : \ q(0) = 0, \ \sum_{i \in I} b_i q_i(t) \geq 1 \right\} .$$

The following is a restatement of Theorem 1-(ii) in terms of a sequence of fluid scaled queues.

*Theorem 2:* For each $n = 1, 2, \ldots$, consider the system under the p-Log scheduling rule in a stationary regime, then, the corresponding sequence of fluid-scaled processes is such that,
$$\limsup_{n \to \infty} \frac{1}{n} \log P\left( \sum_{i \in I} b_i \boldsymbol{q}_i^{(n)}(0) \geq 1 \right) \leq -J_{**} .$$

*Remark 3:* This is the *equivalent* of Theorem 8.4 of [15], and its proof (given in the Appendix) follows the same framework and uses classical Wentzel-Freidlin constructions [18]. The theorem establishes two things: firstly, that the upper bound on the probability of overflow when starting with empty queues, given by Stolyar's refinement of Mogulskii's upper bound, indeed reduces to inf over the cost of GFSPs of interest; and secondly, that a GFSP with the cheapest limiting trajectories $(f, g)$ that can raise the sum queue $\sum_{i \in I} b_i q_i$ to 1, *starting with empty queues*, indeed has a cost arbitrarily close to the cost of the cheapest trajectory *starting in the stationary regime*. See Appendix for a proof.

It is clear that $J_{**} \leq J_*$ (since $-J_*$ was lower bound under any scheduling rule.) To prove the optimality of p-Log rule, we need show $J_{**} = J_*$. In the following section, we develop the results needed to show this; these intermediate steps are summarized in Table I.

## IX. LOCAL FLUID SAMPLE PATH

Let us first motivate the need for defining LFSP (local fluid sample path.) For each $n$, define the set $\mathcal{S}_m^{(n)}$ as the "fluid" scaled version of set $\mathcal{S}_m$ of the state space partition associated with the p-Log rule, i.e.,
$$\mathcal{S}_m^{(n)} = \left\{ x \in \mathbb{R}_+^2 : \ nx \in \mathcal{S}_m \right\} .$$

Then for the $n^{th}$ system, the scheduling decision at time $t$ depends on which set $\mathcal{S}_m^{(n)}$, $m \in \{0, \ldots, k, l+1, \ldots, M'+1\}$ (or the corresponding switching curve) the fluid scaled queue $q^{(n)}(t)$ lies in. As $n \to \infty$, the characteristic function of set $\mathcal{S}_0^{(n)}$ converges pointwise to the characteristic function of $\mathcal{S}_0^{(\infty)} = \{ x \in \mathbb{R}_+^2 : \ x > 0 \}$; while all other scaled sets from the partition collapse to one of the axes. Note that this is true for the partition under any radial sum-rate monotone scheduling rule. Now consider a Lipschitz continuous limiting trajectory $(q, f, g)$ for the fluid scaled process $(q^{(n)}, f^{(n)}, g^{(n)})$. One can show that, if $q(t) \in \mathcal{S}_0^{(\infty)}$, then $\frac{d}{dt} \langle b, q(t) \rangle = \langle b, f'(t) \rangle - \max_{v \in V_{g'(t)}} \langle b, v \rangle$, but if $q(t)$ hits an axis, we lose information about service rates of the queues. Hence, we define a LFSP using a *finer*-than-fluid scaling such that the sets of the partition do not collapse and we are able to state the proper derivative of the limiting queue trajectory.

Consider a GFSP over some interval $[0, T]$ and fix any $\tau \in (0, T)$ such that $q(\tau) \neq 0$. Any sequence $\tau^{(n)} \to \tau$ has a subsequence along which $q^{(n)}(\tau^{(n)}) \to q(\tau)$. Set $\sigma_n = \sqrt{q_*^{(n)}(\tau^{(n)})}/\sqrt{n}$, where, $q_*^{(n)}(\cdot) = \max_{i \in I} q_i^{(n)}(\cdot)$. To obtain a local fluid sample path, we will *magnify* in both space and time the fluid scaled trajectories $(q^{(n)}, f^{(n)}, g^{(n)})$ by a factor of $\sigma_n^{-1}$, i.e., an order $O(\sqrt{n})$ term. More formally, for any fixed $S > 0$, the following re-scaled functions (and their limits mentioned subsequently) over the interval $[\tau^{(n)}, \tau^{(n)} + \sigma_n S]$, parameterized by $s \in [0, S]$, are called the local fluid sample paths[4]: for all $i \in I$ and $m \in \mathcal{M}$,

$$\begin{aligned}
_\diamond q_i^{(n)}(s) &= \frac{1}{\sigma_n} \left( q_i^{(n)}(\tau^{(n)} + \sigma_n s) - q_i^{(n)}(\tau^{(n)}) \right) , \\
_\diamond \hat{q}_i^{(n)}(s) &= \frac{1}{\sigma_n} q_i^{(n)}(\tau^{(n)} + \sigma_n s) , \\
_\diamond d^{(n)}(s) &= {}_\diamond \hat{q}_1^{(n)}(s) - {}_\diamond \hat{q}_2^{(n)}(s) , \\
_\diamond f_i^{(n)}(s) &= \frac{1}{\sigma_n} \left( f_i^{(n)}(\tau^{(n)} + \sigma_n s) - f_i^{(n)}(\tau^{(n)}) \right) , \\
_\diamond g_m^{(n)}(s) &= \frac{1}{\sigma_n} \left( g_m^{(n)}(\tau^{(n)} + \sigma_n s) - g_m^{(n)}(\tau^{(n)}) \right) .
\end{aligned}$$

Then along some subsequence in $n$,
- the functions $({}_\diamond q^{(n)}, {}_\diamond f^{(n)}, {}_\diamond g^{(n)})$ converge uniformly over $[0, S]$ to Lipschitz continuous functions $({}_\diamond q, {}_\diamond f, {}_\diamond g)$;
- for each $i \in I$, the function ${}_\diamond \hat{q}_i^{(n)}$ either converges uniformly over $[0, S]$ to a finite Lipschitz continuous function ${}_\diamond \hat{q}_i$, or is identically equal to $\infty$;
- the function ${}_\diamond d^{(n)}$ converges uniformly over $[0, S]$ to a finite Lipschitz continuous function ${}_\diamond d$, or is identically equal to $+\infty$ or $-\infty$.

We will refer to the point $\left( q(\tau), f(\tau), g(\tau) \right)$ as the *GFSP source point* of the above defined LFSP. Since $q(\tau) \neq 0$, it must be that ${}_\diamond \hat{q}_i(\cdot) = \infty$ for at least one $i \in I$. Note that the local fluid queue, ${}_\diamond q$, is merely a *re-centered* version of ${}_\diamond \hat{q}$, i.e., ${}_\diamond \hat{q}(s) = {}_\diamond q(s) + {}_\diamond \hat{q}(0)$, and is always finite by virtue of this re-centering. Moreover, the trajectory ${}_\diamond q$ dwells in the set $\{ x \in \mathbb{R}^2 | x \geq -{}_\diamond \hat{q}(0) \}$, which is at least a half-plane. Lastly, we have the following relation between the *cost* of LFSP over $[0, S]$ and the refined cost sequence of GFSP over $[\tau^{(n)}, \tau^{(n)} + \sigma_n S]$ (see (9.4) of [15]),

$$J_S({}_\diamond f, {}_\diamond g) - J_0({}_\diamond f, {}_\diamond g) \leq \liminf_{n \to \infty} \frac{1}{\sigma_n} \left( \overline{J}_{\tau^{(n)} + \sigma_n S}^{(n)} - \overline{J}_{\tau^{(n)}}^{(n)} \right) . \tag{16}$$

### A. Scheduling over time scales of LFSP

By (6) the scheduling decision in the interval $[\tau^{(n)}, \tau^{(n)} + \sigma_n S]$ depends on the slope,
$$\frac{h_2 \left( Q \left( n \tau^{(n)} + n \sigma_n s \right) \right)}{h_1 \left( Q \left( n \tau^{(n)} + n \sigma_n s \right) \right)} ,$$

and the sign of ${}_\diamond d^{(n)}(s)$, recall the tie-breaking rule mentioned in the description of p-Log rule in Section IV.

---
[4]The definitions of ${}_\diamond f_i^{(n)}(\cdot)$ and ${}_\diamond g_m^{(n)}(\cdot)$ are the same as in [15], whereas, ${}_\diamond q_i^{(n)}(\cdot)$ are *scaled* as in [15] but *centered* differently.

| Description | Stated in | Relies on |
|---|---|---|
| Under the usual fluid scaling of the queue state space, some sets of the partition collapse (or merge), resulting in loss of information concerning the service rate seen by the queue (e.g., when the queue lies in any of the collapsed sets.) Therefore, the fluid-scaled state space is locally *magnified* enough to recover the merged sets of the partition; these magnified or finer-than-fluid-scaled trajectories are called Local Fluid Sample Paths (LFSP). | Section IX, Lemma 1. | Technique first introduced in [4] |
| Although the vector field $h(\cdot)$ associated with the p-Log rule is not a gradient field, it appears as a gradient field on the state space of finer-than-fluid-scaled queue. A (globally) Lipschitz continuous Lyapunov function is then constructed on this state space. Moreover, under the p-Log scheduling rule, the Lyapunov function is shown to have a strictly negative drift for all LFSP trajectories having low *cost per unit time*. | Lemma 2 | Lemma 1 |
| The strictly negative drift of a Lipschitz continuous Lyapunov function translates into a stronger implication: for all LFSP trajectories with low average cost per unit time over a given time interval, the decrease in the Lyapunov function of the finer-than-fluid-scaled queue must be proportional to the length of the interval. Moreover, a sufficient decrease in the Lyapunov function also implies at least a proportional decrease in the weighted-sum-queue. | Lemma 3 | Lemma 2 |
| The above result is used to show that any fluid-scaled trajectory (GFSP) of interest can be *magnified* to obtain a finer-than-fluid-scaled trajectory (LFSP) such that the cost (per unit increase in weighted sum-queue) of the fluid-scaled trajectory and that of the finer-than-fluid-scaled trajectory are arbitrarily close. | Lemma 4 | Lemma 3 and using technique of Section 11 of [15] |
| Under the p-Log rule, no LFSP exists whose *cost per unit increase in weighted-sum-queue* is strictly less than $J_*$, therefore, the least possible cost under p-Log rule, i.e. $J_{**}$, must be equal to $J_*$ – the upper bound on the cost under any scheduler. | Section 10 | Lemma 4 |

TABLE I
INTERMEDIATE STEPS TOWARDS PROVING THEOREM 1-(III).

Without loss of generality suppose $q_1(\tau) \geq q_2(\tau)$ (and recall that we had $q(\tau) \neq 0$.) Then by (5), for $n$ large enough, we have that,

$$\frac{h_2\left(Q\left(n\tau^{(n)} + n\sigma_n s\right)\right)}{h_1\left(Q\left(n\tau^{(n)} + n\sigma_n s\right)\right)}$$
$$= \frac{b_2}{b_1} \min\left(1, \frac{Q_2\left(n\tau^{(n)} + n\sigma_n s\right)}{\sqrt{Q_1\left(n\tau^{(n)} + n\sigma_n s\right)}}\right),$$
$$= \frac{b_2}{b_1} \min\left(1, \frac{{}_\diamond\hat{q}_2^{(n)}(s)\sqrt{q_1^{(n)}(\tau^{(n)})}}{\sqrt{q_1^{(n)}(\tau^{(n)} + \sigma_n s)}}\right).$$

Then, as $n \to \infty$, the above converges to,

$$\frac{b_2}{b_1} \min\left(1, {}_\diamond q_2(s) + {}_\diamond\hat{q}_2(0)\right),$$

where the convergence is uniform on $[0, S]$. Let us define a vector field ${}_\diamond h$ over the state space of ${}_\diamond q$, i.e., $\{x \in \mathbb{R}^2 | x \geq -{}_\diamond\hat{q}(0)\}$, as follows:

$$\begin{aligned}{}_\diamond h_1(x) &= b_1, \\ {}_\diamond h_2(x) &= b_2 \min\left(1, x_2 + {}_\diamond\hat{q}_2(0)\right).\end{aligned} \quad (17)$$

That is, we can restate the above convergence as,

$$\frac{h_2\left(Q\left(n\tau^{(n)} + n\sigma_n s\right)\right)}{h_1\left(Q\left(n\tau^{(n)} + n\sigma_n s\right)\right)} \to \frac{{}_\diamond h_2({}_\diamond q(s))}{{}_\diamond h_1({}_\diamond q(s))}$$

uniformly on $[0, S]$. Hence, the switching curves on the space of ${}_\diamond q$ are given by,

$$\left\{(x_1, x_2) \in \mathbb{R}^2 : \frac{{}_\diamond h_2(x)}{{}_\diamond h_1(x)} = r_m\right\}$$
$$= \left\{(x_1, x_2) \in \mathbb{R}^2 : x_2 = -{}_\diamond\hat{q}_2(0) + \frac{b_1}{b_2}r_m\right\}$$

for $m \in \{0, \cdots, k\}$, and are now parallel to $x_1$ axis (i.e., the axis of ${}_\diamond q_1(s)$, see Fig. 4)[5]. For each $m \in \{1, \cdots, k\}$, define

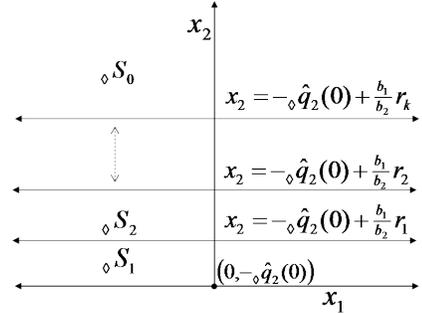

Fig. 4. Partitions and switching curves on space of local fluid queue ${}_\diamond q$.

the set ${}_\diamond\mathcal{S}_m$ by appropriately re-scaling the set $\mathcal{S}_m$, i.e.,

$$\begin{aligned}{}_\diamond\mathcal{S}_m &= \left\{(x_1, x_2) \in \mathbb{R}^2 : r_{m-1} < \frac{{}_\diamond h_2(x)}{{}_\diamond h_1(x)} < r_m\right\}, \\ &= \Big\{(x_1, x_2) \in \mathbb{R}^2 : \\ &\quad -{}_\diamond\hat{q}_2(0) + \frac{b_1}{b_2}r_{m-1} < x_2 < -{}_\diamond\hat{q}_2(0) + \frac{b_1}{b_2}r_m\Big\},\end{aligned} \quad (18)$$

and define,

$$\begin{aligned}{}_\diamond\mathcal{S}_0 &= \left\{(x_1, x_2) \in \mathbb{R}^2 : r_k < \frac{{}_\diamond h_2(x)}{{}_\diamond h_1(x)}\right\}, \\ &= \left\{(x_1, x_2) \in \mathbb{R}^2 : -{}_\diamond\hat{q}_2(0) + \frac{b_1}{b_2}r_k < x_2\right\}.\end{aligned} \quad (19)$$

See Fig. 4 for a graphical illustration of sets ${}_\diamond\mathcal{S}_0, \ldots, {}_\diamond\mathcal{S}_k$. The service rate $\mu(s)$ allocated to the local fluid queue ${}_\diamond q(s)$ at time $s$, depends on which re-scaled set ${}_\diamond\mathcal{S}_m$ (or the associated switching curve) the local fluid queue ${}_\diamond q(s)$ lies in. More formally, the following lemma relates the service rate $\mu(s)$ to the vector field $_\circ h(_\circ q(s))$, and can be derived without much effort from the results shown for the Exponential rule scheduler in [4][6].

---

[5]Recall that we have assumed, without loss of generality, that $q_1(\tau) \geq q_2(\tau)$; had we assumed otherwise, we would have found that the switching curves on the space of ${}_\diamond q$, i.e., $\{x \in \mathbb{R}^2 | x \geq -{}_\diamond\hat{q}(0)\}$ are parallel to the $x_2$ axis (i.e., the axis of ${}_\diamond q_2(s)$). In the sequel, we continue to assume that at the GFSP source point, we have $q_1(\tau) \geq q_2(\tau)$, and therefore, ${}_\diamond h$ is given by (17).

*Lemma 1:* For any LFSP, the following derivatives exist *a.e.* in $[0, S]$ and are finite:

$$\lambda(s) = \frac{d}{ds} {}_\circ f(s) ,$$
$$\gamma(s) = \frac{d}{ds} {}_\circ g(s) ,$$
$$\frac{d}{ds} {}_\circ q(s) = \lambda(s) - \mu(s) ,$$

for some,

$$\mu(s) \in \arg\max_{v \in V_{\gamma(s)}} \langle {}_\circ h({}_\circ q(s)), v \rangle . \qquad (20)$$

*Remark 4:* Note that if $_\circ q_2(s) \geq -_\circ \hat{q}_2(0) + 1$, then $_\circ h(_\circ q(s)) = b$, and the argmax in (20) may not be unique. However, $\mu(s)$ can still be uniquely identified as long as $_\circ d(s) \neq 0$. That is, if $_\circ d(s) > 0$, then $\mu(s)$ is such that $\mu_1(s)$ is the largest possible among all points achieving max in (20); similarly, if $_\circ d(s) < 0$, then $\mu(s)$ is such that $\mu_2(s)$ is the largest. The argmax in (20) may again be non-unique if $_\circ q(s)$ lies on a switching curve, i.e., $_\circ q_2(s) = -_\circ \hat{q}_2(0) + \frac{b_1}{b_2} r_m$ for some $m \in \{0, 1, \cdots, r_k\}$, however, in this case $\mu(s)$ can not be uniquely identified using only the components of LFSP, nor will we need to uniquely identify $\mu(s)$.

We will also need the following two crucial lemmas that show for the p-Log rule what Lemmas 9.2 and 9.3 of [15] show for the Exponential rule. The following two lemmas also implicitly prove the throughput-optimality of p-Log rule, see Appendix for details.

*Lemma 2:* There exist fixed constants $\epsilon_1 > 0$ and $\delta_1 > 0$, and a Lipschitz continuous Lyapunov function $H$ (constructed in the proof below) such that for any regular point $s \in [0, S]$, if

$$\frac{d}{ds} J_s({}_\circ f, {}_\circ g) \leq \epsilon_1 , \quad \text{then} \quad \frac{d}{ds} H({}_\circ q(s)) \leq -\delta_1 .$$

*Proof* If $\frac{d}{ds} J_s({}_\circ f, {}_\circ g)$ is small, then $\lambda(s)$ must be close to the mean arrival rate $\bar{\lambda}$, and $\gamma(s)$ close to the server state distribution $\pi$. Since the capacity region $V_\gamma$ is continuous in the distribution $\gamma$ (see (1)), there exist vectors $\lambda^* < v^*$ such that uniformly on all sufficiently small values of $\epsilon_1$, we have $\lambda(s) < \lambda^*$, and $v^*$ lies in the interior of $V_{\gamma(s)}$. Let $\delta_1 = \min_{i \in I} (v_i^* - \lambda_i^*)$. By Lemma 1, we have,

$$\langle {}_\circ h({}_\circ q(s)), \mu(s) \rangle = \arg\max_{v \in V_{\gamma(s)}} \langle {}_\circ h({}_\circ q(s)), v \rangle ,$$
$$> \langle {}_\circ h({}_\circ q(s)), v^* \rangle . \qquad (21)$$

Note that since $_\circ h_i(x)$ is a function only of $x_i$, the vector field $_\circ h$ is in fact a gradient field associated with a (continuously differentiable) function[7] $H$, i.e., $\nabla H = {}_\circ \hat{h}$ on $\{x \in \mathbb{R}^2 | x \geq -{}_\circ \hat{q}(0)\}$. Moreover, since $_\circ h_i(\cdot)$ is increasing, positive, and bounded, therefore, the function $H(\cdot)$ is convex, increasing in each direction, and Lipschitz continuous. Then we have,

$$\frac{d}{ds} H({}_\circ q(s)) = \langle {}_\circ h({}_\circ q(s)), {}_\circ q'(s) \rangle ,$$
$$= \langle {}_\circ h({}_\circ q(s)), \lambda(s) - \mu(s) \rangle ,$$
$$\leq \langle {}_\circ h({}_\circ q(s)), \lambda^* - v^* \rangle ,$$
$$\leq -\delta_1 ,$$

where the first inequality follows from (21) and the second from the definition of $\delta_1$. ∎

*Lemma 3:* There exist fixed constants $\epsilon_2 > 0$ and $\hat{\delta}_2 > \delta_2 > 0$ such that, if

$$J_S({}_\circ f, {}_\circ g) - J_0({}_\circ f, {}_\circ g) \leq \epsilon_2 S ,$$

then,

$$H({}_\circ q(S)) - H({}_\circ q(0)) \leq -\hat{\delta}_2 S ,$$

which further implies that uniformly for all large $S$, we have the following bound on the change in the weighted sum queue,

$$\langle b, {}_\circ q(S) - {}_\circ q(0) \rangle \leq -\delta_2 S.$$

*Proof* Noting that $H({}_\circ q(s))$ (as a function of $s$) is Lipschitz continuous with some Lipschitz constant denoted by $c$, the proof of first statement is identical to that of Lemma 9.3 of [15]: pick a positive $\epsilon_2 < \epsilon_1$, let $B_1 = \{s \in [0, S] : \frac{d}{ds} J_s({}_\circ f, {}_\circ g) \geq \epsilon_1\}$ and $B_2 = S \setminus B_1$. Then the Lebesgue measures of $B_1$ and $B_2$ satisfy, $\nu(B_1) \leq \frac{\epsilon_2}{\epsilon_1} S$ and $\nu(B_2) \geq (1 - \frac{\epsilon_2}{\epsilon_1}) S$. Finally,

$$H({}_\circ q(S)) - H({}_\circ q(0))$$
$$= \int_{B_1} \frac{d}{ds} H({}_\circ q(s)) + \int_{B_2} \frac{d}{ds} H({}_\circ q(s)) ,$$
$$\leq c \frac{\epsilon_2}{\epsilon_1} S - \delta_1 (1 - \frac{\epsilon_2}{\epsilon_1}) S ,$$
$$= -S \left( \delta_1 - \frac{\epsilon_2}{\epsilon_1} (c + \delta_1) \right) .$$

Fix any positive $\hat{\delta}_2 < \delta_1$, then a sufficiently small $\epsilon_2$ can be chosen to prove the first statement.

To prove the last statement, we proceed as follows. Without loss of generality suppose that the GFSP source point of the LFSP being considered satisfies $q_1(\tau) \geq q_2(\tau)$, and therefore, $_\circ h$ is given by (17). That is, $\nabla_1 H(\cdot) \equiv {}_\circ h_1(\cdot) = b_1$ and $\nabla_2 H(\cdot) \equiv {}_\circ h_2(\cdot) \leq b_2$. Then for any $x \geq y \geq {}_\circ \hat{q}(0)$, we have the upper bound

$$H(x) - H(y) \leq \langle b, x - y \rangle . \qquad (22)$$

If $y_2 \geq -{}_\circ \hat{q}_2(0) + 1$ (see Fig. 4), then for all $z \geq y$, we have $_\circ h(z) = b$ and therefore,

$$H(x) - H(y) = \langle b, x - y \rangle . \qquad (23)$$

If $-{}_\circ \hat{q}_2(0) \leq y_2 < -{}_\circ \hat{q}_2(0) + 1$, then we have the lower bound

$$H(x) - H(y)$$
$$= \int_{y_1}^{x_1} \nabla_1 H(z_1, y_2) \, dz_1 + \int_{y_2}^{x_2} \nabla_2 H(x_1, z_2) \, dz_2 ,$$

---

[6]Specifically, Proposition 1 and results leading from (34) to (35) of [4] relate the service rate seen by the local fluid queue under Exponential rule scheduler to the vector field defining the scheduler.

[7]General conditions for a vector field to form a gradient field can be found in, e.g., [19] (pp. 944-945). In our case, $H(x)$ is simply additive separable, i.e., $H(x) \equiv H_1(x_1) + H_2(x_2)$.

$$\geq b_1(x_1 - y_1) + \int_{-_\diamond\hat{q}_2(0)+1}^{x_2} b_2 \, dz_2 \;,$$
$$= b_1(x_1 - y_1) + \int_{y_2}^{x_2} b_2 \, dz_2 - \int_{y_2}^{-_\diamond\hat{q}_2(0)+1} b_2 \, dz_2 \;,$$
$$\geq \langle b, x-y \rangle - b_2 \;. \quad (24)$$

Combining (22–24), we have the following bounds for any $x \geq y \geq _\diamond\hat{q}(0)$,

$$\langle b, x-y \rangle - (b_1+b_2) \leq H(x) - H(y) \leq \langle b, x-y \rangle \;.$$

Using this we get,

$$\begin{aligned} -\hat{\delta}_2 S &\geq H\big(_\diamond q(S)\big) - H\big(_\diamond q(0)\big) \;,\\ &\geq \langle b, _\diamond q(S) - _\diamond q(0)\rangle - (b_1+b_2)\\ \Rightarrow -\hat{\delta}_2 S \left(1 - \frac{b_1+b_2}{S}\right) &\geq \langle b, _\diamond q(S) - _\diamond q(0)\rangle \;. \end{aligned}$$

Now one can take $S$ large enough to obtain a $\delta_2$. ∎

Next, using Lemma 3 above and identical to the result in Section 11 of [15], we show that any GFSP that raises the weighted sum queue to unity contains a LFSP with cost *close* to that of the GFSP. Subsequently, we will use this lemma on a GFSP of cost close to $J_{**}$ (see (15)) in order to obtain a LFSP with cost close to $J_{**}$.

*Lemma 4:* Suppose a GFSP $\psi$ is given that satisfies $q(0)=0$, $\langle b, q(T)\rangle = 1$ for some $T>0$ and has a cost $\overline{J}_T < \infty$. Then, for an arbitrarily small $\epsilon > 0$, an LFSP $(_\diamond q, _\diamond\hat{q}, _\diamond d, _\diamond f, _\diamond g)$ over an arbitrarily large interval $[0, S]$ can be constructed from the elements of $\psi$, such that,

$$\langle b, _\diamond q(S) - _\diamond q(0) \rangle \geq \theta S \;, \quad (25)$$

for some $\theta > 0$ (independent of $\epsilon$), and cost (per unit increase in weighted sum queue) of this LFSP is bounded above by $\overline{J}_T + \epsilon$, i.e.,

$$\frac{J_S(_\diamond f, _\diamond g) - J_0(_\diamond f, _\diamond g)}{\langle b, _\diamond q(S) - _\diamond q(0)\rangle} \leq \overline{J}_T + \epsilon \;. \quad (26)$$

*Proof* Components of the given GFSP $\psi$ satisfy,

$$\frac{(\overline{J}_T - \overline{J}_0)}{\langle b, q(T) - q(0)\rangle} = \overline{J}_T \;. \quad (27)$$

For any $0 < \xi_1 < \xi_2 < 1$, define time $t_1 = \max(t : \langle b, q(t)\rangle = \xi_1)$ and time $t_2 = \min(t > t_1 : \langle b, q(t)\rangle = \xi_2)$. Then for any $\epsilon > 0$, there must exist $0 < \xi_1 < \xi_2 < 1$ such that,

$$\frac{\overline{J}_{t_2} - \overline{J}_{t_1}}{\langle b, q(t_2) - q(t_1)\rangle} < \overline{J}_T + \frac{\epsilon}{2} \;, \quad (28)$$

or (27) cannot hold (Dirichlet's box principle). Fix an $S > 0$ large enough as required by Lemma 3. Then there exists a sequence $\{\tau^{(n)}\}$ within $[t_1, t_2]$ such that,

$$\left\langle b, q^{(n)}(\tau^{(n)} + \sigma_n S) - q^{(n)}(\tau^{(n)})\right\rangle > 0 \;,$$

$$\frac{\overline{J}^{(n)}_{(\tau^{(n)}+\sigma_n S)} - \overline{J}^{(n)}_{\tau^{(n)}}}{\langle b, q^{(n)}(\tau^{(n)} + \sigma_n S) - q^{(n)}(\tau^{(n)})\rangle} < \overline{J}_T + \epsilon \;,$$

where, as before, $\sigma_n = \sqrt{q_*^{(n)}(\tau^{(n)})}/\sqrt{n}$; such a sequence must exist, otherwise (28) cannot hold (another application of Dirichlet's box principle, and consequence of the convergence of $\overline{J}^{(n)}$ and $q^{(n)}$). The above two inequalities can be re-written as,

$$\left\langle b, _\diamond q^{(n)}(S) - _\diamond q^{(n)}(0)\right\rangle > 0 \;, \quad (29)$$

$$\frac{\sigma_n^{-1}\big(\overline{J}^{(n)}_{(\tau^{(n)}+\sigma_n S)} - \overline{J}^{(n)}_{\tau^{(n)}}\big)}{\langle b, _\diamond q^{(n)}(S) - _\diamond q^{(n)}(0)\rangle} < \overline{J}_T + \epsilon \;. \quad (30)$$

We also have that $q(\cdot) \neq 0$ over interval $[t_1, t_2]$. Now we can pick a subsequence in $n$ along which $\tau^{(n)} \to \tau \in [t_1, t_2]$ and $\big(_\diamond q^{(n)}, _\diamond\hat{q}^{(n)}, _\diamond d^{(n)}, _\diamond f^{(n)}, _\diamond g^{(n)}\big)$ converge to $\big(_\diamond q, _\diamond\hat{q}, _\diamond d, _\diamond f, _\diamond g\big)$, as described in Section IX, thus obtaining a LFSP (with $\big(q(\tau), f(\tau), g(\tau)\big)$ being its GFSP source point.) From (29) we also have that $\langle b, _\diamond q(S) - _\diamond q(0)\rangle \geq 0$. Then by Lemma 3, we get,

$$J_S(_\diamond f, _\diamond g) - J_0(_\diamond f, _\diamond g) > \epsilon_2 S \;,$$

which, alongside (30) and (16), further implies that for some $\theta > 0$,

$$\langle b, _\diamond q(S) - _\diamond q(0)\rangle \geq \theta S \;,$$

and finally,

$$\frac{J_S(_\diamond f, _\diamond g) - J_0(_\diamond f, _\diamond g)}{\langle b, _\diamond q(S)\rangle - \langle b, _\diamond q(0)\rangle} \leq \overline{J}_T + \epsilon \;.$$

∎

## X. PROOF OF THEOREM 1-(III): OPTIMALITY OF THE P-LOG RULE

Recall that to prove optimality of p-Log rule (Theorem 1-(iii)), we need show $J_{**} \geq J_*$. We will do this by showing that assuming $J_{**} < J_*$ leads to a contradiction with the definition of $J_*$.

Suppose $J_{**} < J_*$, then by definition of $J_{**}$ in (15), there exists a GFSP $\psi$ satisfying $q(0)=0$, $\langle b, q(T)\rangle = 1$ for some finite $T>0$, and having a cost $\overline{J}_T < J_*$. Then by Lemma 4, from the components of GFSP $\psi$, we can construct an LFSP $\big(_\diamond q, _\diamond\hat{q}, _\diamond d, _\diamond f, _\diamond g\big)$ satisfying (25) and (26) for an arbitrarily large $S$ and an $\epsilon$ small enough so that $J_{***} \equiv \overline{J}_T + \epsilon < J_*$.

Since $_\diamond\hat{q}_i(\cdot) = \infty$ for at least one $i \in I$, without loss of generality suppose $_\diamond\hat{q}_1(\cdot) = \infty$, thus all switching curves on the space of $_\diamond q$ are parallel to $_\diamond q_1$ axis. Recall that the lower boundary of the set $_\diamond\mathcal{S}_0$ is given by the switching curve $x_2 = -_\diamond\hat{q}_2(0) + \frac{b_1}{b_2}r_k$ (see Fig. 4). Let $S_1$ and $S_2$ respectively be the first and the last time in $[0, S]$ such that the trajectory $_\diamond q_2(s) \leq -_\diamond\hat{q}_2(0) + \frac{b_1}{b_2}r_k$, with $S_1 = S_2 = S$ if $_\diamond q_2(s)$ never hits $[-_\diamond\hat{q}_2(0), -_\diamond\hat{q}_2(0) + \frac{b_1}{b_2}r_k]$. Note that the trajectory of $_\diamond q(s)$ lies in $_\diamond\mathcal{S}_0$ in $(0, S_1)$ and $(S_2, S)$. Then one of the following must be true over the interval $[0, S_1]$ (similarly $[S_2, S]$):

(a) $\langle b, _\diamond q(0)\rangle < \langle b, _\diamond q(S_1)\rangle$ and the cost per unit increase in weighted-sum-queue over the interval $[0, S_1]$ is less than $J_{***}$, i.e.

$$\frac{J_{S_1}(_\diamond f, _\diamond g) - J_0(_\diamond f, _\diamond g)}{\langle b, _\diamond q(S_1) - _\diamond q(0)\rangle} \leq J_{***} \;.$$

(b) $\langle b, {}_\diamond q(0)\rangle < \langle b, {}_\diamond q(S_1)\rangle$ and the cost per unit increase in sum-queue over the interval $[0, S_1]$ is strictly greater than $J_{***}$, i.e.

$$\frac{J_{S_1}({}_\diamond f, {}_\diamond g) - J_0({}_\diamond f, {}_\diamond g)}{\langle b, {}_\diamond q(S_1) - {}_\diamond q(0)\rangle} > J_{***} \ .$$

(c) $\langle b, {}_\diamond q(0)\rangle \geq \langle b, {}_\diamond q(S_1)\rangle$.

If (a) is true for either one of the intervals (suppose its true for $[0, S_1]$,) we proceed as follows: define vectors $\hat{\gamma}$, $\hat{\lambda}$, and $\hat{\mu}$ as the *average* server state distribution, arrival rate, and service rate respectively over $[0, S_1]$, i.e.,

$$\left(\hat{\lambda}, \hat{\gamma}, \hat{\mu}\right) = \frac{1}{S_1} \int_0^{S_1} \left(f'(s), g'(s), \mu(s)\right) ds.$$

By Lemma 1 and the fact that $({}_\diamond q(s) : 0 < s < S_1)$ lies in ${}_\diamond \mathcal{S}_0$, we have $\mu(s) \in \arg\max_{v \in V_{\gamma(s)}} \langle b, v\rangle$. This and the *linearity* of $\mu(s)$ in $\gamma(s)$ (see (1)) implies $\hat{\mu} \in \arg\max_{v \in V_{\hat{\gamma}}} \langle b, v\rangle$. Then,

$$\begin{aligned}\langle b, {}_\diamond q(S_1) - q(0)\rangle &= \left\langle b, \hat{\lambda} - \hat{\mu}\right\rangle S_1 \ , \\ &\leq \left\langle b, \hat{\lambda} - v^*(\hat{\lambda}, \hat{\gamma})\right\rangle S_1 \ , \end{aligned} \quad (31)$$

where $v^*(\hat{\lambda}, \hat{\gamma})$ is as defined in (8). Finally,

$$\begin{aligned}J_* > J_{***} &\geq \frac{J_{S_1}({}_\diamond f, {}_\diamond g) - J_0({}_\diamond f, {}_\diamond g)}{\langle b, {}_\diamond q(S_1) - {}_\diamond q(0)\rangle} \ , \\ &\geq \frac{\left(L_{(f)}(\hat{\lambda}) + L_{(g)}(\hat{\gamma})\right) S_1}{\langle b, {}_\diamond q(S_1) - {}_\diamond q(0)\rangle} \ , \\ &\geq \frac{L_{(f)}(\hat{\lambda}) + L_{(g)}(\hat{\gamma})}{\left\langle b, \hat{\lambda} - v^*(\hat{\lambda}, \hat{\gamma})\right\rangle} \ , \end{aligned}$$

where the first inequality follows from the assumption that (a) is true, the second from convexity of rate functions, and the last one from (31). However, by the definition of $J_*$ in (9), the right side of last inequality cannot be less than $J_*$, giving the contradiction we needed; therefore, we must have $J_* = J_{**}$.

Now, if (a) is not true for both intervals $[0, S_1]$ and $[S_2, S]$, then we proceed as follows. Recall that our LFSP satisfies (25) and (26), i.e.,

$$\langle b, {}_\diamond q(S) - {}_\diamond q(0)\rangle \geq \theta S > 0 \ , \quad (32)$$

and,

$$\begin{aligned}J_{***} &\geq \frac{J_S({}_\diamond f, {}_\diamond g) - J_0({}_\diamond f, {}_\diamond g)}{\langle b, {}_\diamond q(S) - {}_\diamond q(0)\rangle}, \\ &= \frac{J_{S_1}({}_\diamond f, {}_\diamond g) - J_0({}_\diamond f, {}_\diamond g)}{\langle b, {}_\diamond q(S_1) - {}_\diamond q(0)\rangle} \times \frac{\langle b, {}_\diamond q(S_1) - {}_\diamond q(0)\rangle}{\langle b, {}_\diamond q(S) - {}_\diamond q(0)\rangle} + \\ & \quad \frac{J_{S_2}({}_\diamond f, {}_\diamond g) - J_{S_1}({}_\diamond f, {}_\diamond g)}{\langle b, {}_\diamond q(S_2) - {}_\diamond q(S_1)\rangle} \times \frac{\langle b, {}_\diamond q(S_2) - {}_\diamond q(S_1)\rangle}{\langle b, {}_\diamond q(S) - {}_\diamond q(0)\rangle} + \\ & \quad \frac{J_S({}_\diamond f, {}_\diamond g) - J_{S_2}({}_\diamond f, {}_\diamond g)}{\langle b, {}_\diamond q(S) - {}_\diamond q(S_2)\rangle} \times \frac{\langle b, {}_\diamond q(S) - {}_\diamond q(S_2)\rangle}{\langle b, {}_\diamond q(S) - {}_\diamond q(0)\rangle} \ . \end{aligned} \quad (33)$$

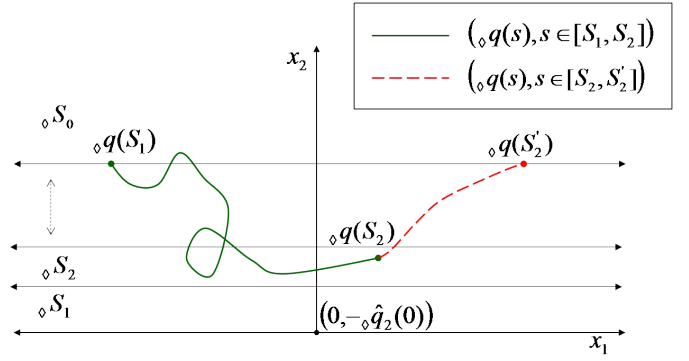

Fig. 5. Illustration of original trajectory $({}_\diamond q(s), s \in [S_1, S_2])$ and the extension $({}_\diamond q(s), s \in [S_2, S'_2])$ in order to obtain ${}_\diamond q_2(S'_2) = {}_\diamond q_2(S_1)$.

Therefore, if (a) is not true for both intervals $[0, S_1]$ and $[S_2, S]$ (equivalently, (b) and/or (c) are true over these intervals), then for (33) to hold, we must have,

$$J_{***} \geq \frac{J_{S_2}({}_\diamond f, {}_\diamond g) - J_{S_1}({}_\diamond f, {}_\diamond g)}{\langle b, {}_\diamond q(S_2) - {}_\diamond q(S_1)\rangle} \ , \quad (34)$$

and, for some fixed $\theta_1 > 0$,

$$\langle b, {}_\diamond q(S_2) - {}_\diamond q(S_1)\rangle \geq \theta_1 \langle b, {}_\diamond q(S) - {}_\diamond q(0)\rangle \ .$$

The above, along with (32) gives,

$$\langle b, {}_\diamond q(S_2) - {}_\diamond q(S_1)\rangle \geq \theta_1 \theta S \ . \quad (35)$$

Moreover, since $\langle b, {}_\diamond q(\cdot)\rangle$ is Lipschitz, for some fixed $\theta_2 > 0$,

$$S_2 - S_1 \geq \theta_2 \langle b, {}_\diamond q(S_2) - {}_\diamond q(S_1)\rangle \geq \theta_2 \theta_1 \theta S \ .$$

This, along with (35) and Lemma 3, imply,

$$\begin{aligned}J_{S_2}({}_\diamond f, {}_\diamond g) - J_{S_1}({}_\diamond f, {}_\diamond g) &\geq \epsilon_2 (S_2 - S_1) \ , \\ &\geq \epsilon_2 \theta_2 \theta_1 \theta S \ . \end{aligned} \quad (36)$$

Subsequently, we will use (35) and (36) to make the quantities on the left side of these inequalities as large as needed by choosing a large $S$.

Consider the trajectory $({}_\diamond q(s), s \in [S_1, S_2])$ which is an element of the LFSP $({}_\diamond q, {}_\diamond \hat{q}, {}_\diamond d, {}_\diamond f, {}_\diamond g)$ over $[S_1, S_2]$. For some $S'_2 \geq S_2$, we can append to this LFSP an *extension* over the time interval $[S_2, S'_2]$ so as to obtain ${}_\diamond q_2(S'_2) = {}_\diamond q_2(S_1)$ (see Fig. 5). Moreover, since the terminal values ${}_\diamond q_2(S_1)$ and ${}_\diamond q_2(S_2)$, lie within the bounded interval $[-{}_\diamond \hat{q}_2(0), -{}_\diamond \hat{q}_2(0) + \frac{b_1}{b_2} r_k]$, therefore the extension LFSP can be constructed such that it has a bounded cost, i.e.,

$$J_{S'_2}({}_\diamond f, {}_\diamond g) - J_{S_2}({}_\diamond f, {}_\diamond g) \leq \Delta \ , \quad (37)$$

and a bounded increase in weighted-sum-queue, i.e.,

$$|\langle b, {}_\diamond q(S'_2) - {}_\diamond q(S_2)\rangle| \leq \Delta \ , \quad (38)$$

for some large fixed $\Delta < \infty$ which is independent of any components of the LFSP over $[0, S]$, (e.g. the value of $S$ or the terminal values ${}_\diamond q_2(S_1)$ and ${}_\diamond q_2(S_2)$.) Finally, the constant $S$ can be chosen large enough such that, by (35) and (38), we have,

$$\langle b, {}_\diamond q(S'_2) - {}_\diamond q(S_1)\rangle > 0 \ ,$$

and by (34)–(38), for some small $\epsilon_4 > 0$ that satisfies $J_{***} + \epsilon_4 < J_*$, we have,

$$\frac{J_{S_2'}(\diamond f, \diamond g) - J_{S_1}(\diamond f, \diamond g)}{\langle b, \diamond q(S_2') \rangle - \langle b, \diamond q(S_1) \rangle} \leq J_{***} + \epsilon_4 < J_* . \quad (39)$$

Set $c_0 = -{\diamond}\hat{q}_2(0)$ and for each $m \in \{1, \cdots, k\}$, choose a $c_m \in \left(-{\diamond}\hat{q}_2(s) + \frac{b_1}{b_2}r_{m-1}, -{\diamond}\hat{q}_2(s) + \frac{b_1}{b_2}r_m\right)$ such that the counting measure of set $\{s \in [0, S_2'] : \diamond q(s) = c_m\}$ is finite; such $\{c_m\}$ exist since $\diamond q$ is Lipschitz. Lastly, choose a $c_{k+1} < \infty$ large enough such that $\max_{s \in [S_1, S_2']} \diamond q_2(s) < c_{k+1}$. For each $m \in \{1, \cdots, k+1\}$, define a set $C_m = \{s \in [S_1, S_2'] : \diamond q_2(s) \in [c_{m-1}, c_m]\}$. We make the following three observations which will be used in the sequel: for each $m \in \{1, \cdots, k+1\}$,

(i) the trajectory $(\diamond q_2(s), s \in [S_1, S_2'])$ intersects with end points of interval $[c_{m-1}, c_m]$ only finitely many times, therefore, the corresponding set $C_m$ can be written as a union of finitely many intervals;

(ii) if set $C_m$ is non-empty, then
$$\diamond q_2(\min_{s \in C_m} s) = \diamond q_2(\max_{s \in C_m} s);$$

(iii) the trajectory $(\diamond q(s), s \in C_m)$ can intersect with at most one switching curve $x_2 = -{\diamond}\hat{q}_2(s) + \frac{b_1}{b_2}r_{m-1}$, or equivalently, with at most two adjacent regions $\diamond S_{(\cdot)}$, (e.g., for $m = k+1$ the above trajectory can intersect with adjacent regions $\diamond S_k$ and $\diamond S_0$).

Then using (i) and (ii) above, for all sets $C_m$, we must have,

$$\int_{C_m} \diamond q_2'(s) \, ds = 0 . \quad (40)$$

Moreover, there exists a set $C_m$ such that,

$$\int_{C_m} \langle b, \diamond q'(s) \rangle \, ds > 0 ,$$

and,

$$\frac{\int_{C_m} \left(L_{(f)}(\diamond f'(s)) + L_{(g)}(\diamond g'(s))\right) ds}{\int_{C_m} \langle b, \diamond q'(s) \rangle \, ds} < J_* ,$$

otherwise (39) will not hold. With a set $C_m$ for which the above two hold, let,

$$\left(\hat{\lambda}, \hat{\gamma}, \hat{\mu}\right) = \frac{1}{\nu(C_m)} \int_{C_m} \left(\diamond f'(s), \diamond g'(s), \mu(s)\right) ds .$$

By (iii) above, (18–20), and the fact that $\diamond d(s) > 0$ in $[S_1, S_2']$, the service vector $\hat{\mu}$ is a convex combination of at most two adjacent vertices of capacity region $V_{\hat{\gamma}}$ and lies on the *facet* with outer normal slope $r_{m-1}$. That is, $\hat{\mu}$ is necessarily a maximal element of $V_{\hat{\gamma}}$. This, together with the fact that $\hat{\lambda}_2 = \hat{\mu}_2$ which follows from (40), gives $\hat{\mu} = \arg\min_{v \in V_{\hat{\gamma}}} \left\langle b, (\hat{\lambda} - v)^+ \right\rangle = v^*(\hat{\lambda}, \hat{\gamma})$, and then,

$$\int_{C_m} \langle b, \diamond q'(s) \rangle \, ds = \left\langle b, \hat{\lambda} - v^*(\hat{\lambda}, \hat{\gamma}) \right\rangle \nu(C_m) .$$

Finally,

$$\begin{aligned} J_* &> \frac{\int_{C_m} \left(L_{(f)}(\diamond f'(s)) + L_{(g)}(\diamond g'(s))\right) ds}{\int_{C_m} \langle b, \diamond q'(s) \rangle \, ds} , \\ &\geq \frac{L_{(f)}(\hat{\lambda}) + L_{(g)}(\hat{\gamma})}{\left\langle b, \hat{\lambda} - v^*(\hat{\lambda}, \hat{\gamma}) \right\rangle} . \end{aligned}$$

By the definition of $J_*$, the right side of last inequality cannot be less than $J_*$, giving the required contradiction; hence we conclude that $J_* = J_{**}$. ∎

## XI. CONCLUSION AND EXTENSIONS

In order to minimize the asymptotic probability of weighted-sum-queue overflow, the desirable mode of overflow is one where queues may build up at different rates, however, the total weighted service rate seen by the queues is the highest possible, (service rate subject to being not more than the arrival rate.) The p-Log scheduling rule minimizes the asymptotic probability of weighted-sum-queue overflow and exhibits such a mode of overflow. This property of p-Log rule is related to the collapse under fluid scaling of all but one set of the state space partition to either of the axes; the set that does not collapse is the horn-shaped weighted-max-sum-rate set. This collapse under fluid scaling is typical of the partition under any radial sum-rate monotone scheduler. However, the scaling or magnifying factor required to obtain a useful LFSP, and the shape of the sets of partition on the local fluid state space will vary for different radial sum-rate monotone schedulers. In this regard, p-Log rule yields an easily tractable partition of the local fluid space where all switching curves are parallel to one of the axes.

Recently in [26], the authors have reported a promising framework to relate the gradient field associated with a MaxWeight-type scheduler to the modes of overflow and large deviations of the appropriately scaled queue process. They are able to show that the Log rule indeed minimizes the asymptotic probability of sum-queue overflow; the framework, however, does not cover the p-Log or Exp rule since the vector field associated with either of these schedulers is not a gradient field. In this regard, Lemmas 2 and 3 of this paper suggest that it may be possible in some cases to *locally* replace the vector field with a gradient field and thus obtain a suitable Lyapunov function, and relate the negative drift of this Lyapunov function to that of the quantity of interest (weighted sum-queue in our case).

The following extensions of the main results of the paper as well as the system model are possible without much effort.

First, the lower bound (i.e., Theorem 1-i and its proof) goes through without any changes for any fixed number of queues (instead of only two) sharing the time-varying server.

Second, the main result (Theorem 1) is also applicable to the following different and *simpler* system model. Instead of a single server with time-varying state $m(t) \in \{1, 2, \ldots, M\}$ having distribution $\pi$, suppose there are $M$ distinct servers with fixed but asymmetric capacities across the two queues; such servers are typically called parallel "unrelated" machines (see, e.g., [27]). More specifically, the $m^{th}$ server, if allocated to the $i^{th} \in I$ queue, can serve $\pi_i \mu_i^m \in \mathbb{Z}_+$ packets from the queue. The total service offered to a queue is taken to be the sum of service offered by each server assigned to that queue. Then the scheduling problem is to dynamically assign the servers to the queues based on the queue state. When queue is in state $Q$, the p-Log scheduler in this context allocates the $m^{th}$ server to the queue $i_{pLog}^*(Q,m)$ as given by (6). This system model is simpler in that the only random process now driving the system are the arrivals, but is also different from the original model in that there are multiple parallel servers with each server having asymmetric capacities across the two queues.

Third, the main result also goes through if the capacity regions $V^m$ are permitted to be arbitrary convex polyhedra instead of just triangles obtained as a convex hull of service vectors $(0,0), (\mu_1^m, 0)$, and $(0, \mu_2^m)$. That is, in a more general model, in any server state $m$, the server can be permitted to operate at any one of the $k_m$ service vectors from the set,

$$\left\{ \left(\mu_1^m(1), \mu_2^m(1)\right), \cdots, \left(\mu_1^m(k_m), \mu_2^m(k_m)\right) \right\} .$$

The region $V^m$ then will be the convex hull of the $k_m$ vertices associated with state $m$. The only change needed is to generalize the definition of p-Log rule as follows. When the system is in state $(Q,m)$, operate the server at a service vector $\mu_{pLog}^*(Q,m) \in V^m$ given by,

$$\mu_{pLog}^*(Q,m) \in \arg\max_{y \in V^m} \langle y, h(Q) \rangle ,$$

where, in the case of a tie, if $Q_1 \geq Q_2$, then $\mu *_{pLog}(Q,m)$ maps to the maximizer with the largest capacity for Queue 1, otherwise $\mu_{pLog}^*(Q,m)$ maps to the maximizer with the largest capacity for Queue 2. This generalization affects the fluid and local fluid sample paths through (20) in Lemma 1, which can be shown to hold exactly as in [20], [21].

Besides the above extensions, we will conclude by stating one more interesting application of the p-Log scheduler. A throughput-optimal scheduler can also be used to offer minimum and maximum average service rate guarantees to *infinitely backlogged* queues, referred to as tasks, sharing a time-varying server/wireless channel [22], [23]. This is done by using virtual token queues that are fed by deterministic arrivals at a constant rate $\lambda_i$, and making scheduling decisions to serve tasks based on the virtual token queues (augmented with a scheduling rule to use when all token queues are empty). If rates $\lambda_i$ are feasible (i.e., vector $\lambda$ lies in the interior of capacity region $V_\pi$ associated with the time-varying server), then under any throughput-optimal scheduler, each task $i$ will be offered an average service rate $v_i \geq \lambda_i$ (such that $v \in V_\pi$). However, if rates $\lambda_i$ are not feasible, then main result of this paper implies that the average service rate vector $v$ has the following interesting and desirable property under p-Log rule: $\langle b, v \rangle$ is maximized subject to $v_i \leq \lambda_i$. That is, p-Log rule splits the tasks in two sets, for one set of tasks $v_i = \lambda_i$, whereas for the other $v_i < \lambda_i$, and the sets are chosen such that the total weighted service rate $\langle b, v \rangle$ is maximized.

## APPENDIX

We begin by developing the necessary results need for the proof of Theorem 2. The corresponding proof in [15] has some parts– lemmas and theorems– that are not specific to the Exp rule; these mostly go through by interpreting $q_*(t)$ (which is the notation for max-queue in [15]) as weighted-sum-queue $\sum_{i \in I} b_i q_i(t)$, while other lemmas and theorems require more work specific to the p-Log rule. We proceed by stating the following two theorems whose proofs are short and identical to those of Theorem 8.5 and Theorem 8.6 in [15] by interpreting $q_*(t)$ as $\sum_{i \in I} b_i q_i(t)$.

*Theorem 3:* (See Theorem 8.5 of [15]) For any fixed $T \geq 0$ and $0 \leq c < 1$, let us denote

$$J_{**, \leq T, c} = \inf_\psi \left\{ \overline{J}_t | \psi : \sum_{i \in I} b_i q_i(0) \leq c \text{ and } \right.$$
$$\left. \sum_{i \in I} b_i q_i(t) \geq 1 \text{ for some } t \leq T \right\}.$$

Then, we have,

$$\limsup_{n \to \infty} \frac{1}{n} \log \sup_{\sum_{i \in I} b_i q_i^{(n)}(0) \leq c} P\left( \sup_{t \in [0,T]} \sum_{i \in I} b_i \boldsymbol{q}_i^{(n)}(t) > 1 \right)$$
$$\leq -J_{**, \leq T, c} ,$$

and as $c \to 0$,

$$J_{**, \leq T, c} \nearrow J_{**, \leq T, 0} = \inf_{t \leq T} J_{**, t} . \quad (41)$$

*Theorem 4:* (See Theorem 8.6 of [15]) For any fixed $C^* > \delta > 0$, and $T > 0$, let us denote

$$K(C^*, \delta, T) = \inf_\psi \left\{ \overline{J}_T | \psi : \sum_{i \in I} b_i q_i(0) \leq C^* \text{ and } \right.$$
$$\left. \sum_{i \in I} b_i q_i(t) \geq \delta \text{ for all } t \in [0,T] \right\} .$$

Then, we have,

$$\limsup_{n \to \infty} \frac{1}{n} \log \sup_{\sum_{i \in I} b_i q_i^{(n)}(0) \leq C^*} P\left( \inf_{t \in [0,T]} \sum_{i \in I} b_i \boldsymbol{q}_i^{(n)}(t) \geq \delta \right)$$
$$\leq -K(C^*, \delta, T) .$$

*Theorem 5:* (See Theorem 8.7 of [15]) For any $C^* > 0$, there exists $\Delta_1 > 0$ such that for all sufficiently large $T$ and all $\delta \in (0, C^*)$, we have $K(C^*, \delta, T) \geq \Delta_1 T$.

*Proof* It is clear that $K(C^*, \delta, T)$ is increasing in $\delta$. We will show that there exists an $\epsilon_5 > 0$ and $\delta_5 > 0$ such that for any GFSP satisfying $\langle b, q(\cdot) \rangle > 0$ over $[0, T]$, if $\overline{J}_T - \overline{J}_0 \leq \epsilon_5 T$, then $\langle b, q(T) - q(0) \rangle \leq -\delta_5 T$, (42) hence for all $T > C^*/\delta_5$ and all $\delta \in (0, C^*)$, we must have $K(C^*, \delta, T) > \epsilon_5 T$, thus the desired result.

Fix $S$ large enough as required by Lemma 3 and recall $\epsilon_2$ and $\delta_2$ therein. For each $n$, define sequence $\{\tau_l^{(n)}, \, l = 0, 1, ..., l^*\}$ in interval $[0, T]$ such that $\tau_0^{(n)} = 0$, and,

$$\tau_{l+1}^{(n)} = \tau_l^{(n)} + \sigma_n(\tau_l^{(n)})S ,$$

where $\sigma_n(\cdot) = \sqrt{q_*^{(n)}(\cdot)}/\sqrt{n}$, and $l^*$ is the largest integer (that depends on $n$) such that $\tau_{l^*}^{(n)} \leq T$. Let $B_2^{(n)}$ be the union of intervals $[\tau_l^{(n)}, \tau_{l+1}^{(n)}]$ over which the refined cost is strictly less than $\epsilon_2 \sigma_n(\tau_l^{(n)})S$, i.e.,

$$B_2^{(n)} = \{\cup [\tau_l^{(n)}, \tau_{l+1}^{(n)}]: \overline{J}_{\tau_{l+1}^{(n)}}^{(n)} - \overline{J}_{\tau_l^{(n)}}^{(n)} < \epsilon_2 \sigma_n(\tau_l^{(n)})S\},$$

and $B_1^{(n)} = [0,T] \setminus B_2^{(n)}$. Now pick a positive $\epsilon_5 < \epsilon_2/2$. From this point on, we assume $n$ large enough. The Lebesgue measures of $B_1^{(n)}$ and $B_2^{(n)}$ satisfy $\nu(B_1^{(n)}) \leq \frac{2\epsilon_5}{\epsilon_2}T$ and $\nu(B_2^{(n)}) \geq (1 - \frac{2\epsilon_5}{\epsilon_2})T$. Moreover, over any interval $[\tau_l^{(n)}, \tau_{l+1}^{(n)}] \subseteq B_2^{(n)}$, we must have,

$$\langle b, q^{(n)}(\tau_{l+1}^{(n)}) - q^{(n)}(\tau_l^{(n)}) \rangle \leq -\frac{\delta_2}{2}\sigma_n(\tau_l^{(n)})S.$$

otherwise we could construct a LFSP contradicting Lemma 3 by choosing a subsequence in $n$ along which the above inequality does not hold and $\tau_l^{(n)}$ converges to some $\tau \in [0,T]$. Then we have the following bound on the total increment of $\langle b, q^{(n)}(\cdot) \rangle$ over $B_2^{(n)}$,

$$\sum_{[\tau_l^{(n)}, \tau_{l+1}^{(n)}] \subseteq B_2^{(n)}} \langle b, q^{(n)}(\tau_{l+1}^{(n)}) - q^{(n)}(\tau_l^{(n)}) \rangle$$
$$\leq -\frac{\delta_2}{2}(1 - \frac{2\epsilon_5}{\epsilon_2})T.$$

Also, because we have assumed bounded arrivals, there is a finite $c_1 > 0$ such that the total increment of $\langle b, q^{(n)}(\cdot) \rangle$ over $[0,T] \setminus B_2^{(n)}$ satisfies,

$$\langle b, q^{(n)}(T) - q^{(n)}(\tau_{l^*}) \rangle +$$
$$\sum_{[\tau_l^{(n)}, \tau_{l+1}^{(n)}] \subseteq B_1^{(n)}} \langle b, q^{(n)}(\tau_{l+1}^{(n)}) - q^{(n)}(\tau_l^{(n)}) \rangle \leq c_1 \frac{2\epsilon_5}{\epsilon_2}T.$$

Finally, summing the above two inequalities and taking limit, we get,

$$\langle b, q(T) - q(0) \rangle \leq \limsup_{n \to \infty} \langle b, q^{(n)}(T) - q^{(n)}(0) \rangle,$$
$$\leq -T\left(\frac{\delta_2}{2} - \frac{\epsilon_5}{\epsilon_2}(\delta_2 + 2c_1)\right).$$

Now we can fix a positive $\delta_5 < \delta_2/2$ and choose $\epsilon_5 > 0$ small enough to satisfy (42), completing the proof. ∎

*Remark 5:* While the reader may conclude the throughput-optimality of the p-Log rule from Lemma 2 alone, nevertheless, the throughput-optimality explicitly bears out as a corollary of the above theorem. Specifically, (42) shows that for any zero cost limiting trajectories $(f, g, q)$– i.e. $(f^{(n)}, g^{(n)}, q^{(n)})$ converge u.o.c to $(f, g, q)$ with probability 1– with initial condition $\langle b, q(0) \rangle = 1$, we have $\langle b, q(T) \rangle = 0$ for all $T > \frac{1}{\delta_5}$. That is, for any deterministic initial state satisfying $\langle b, q^{(n)}(0) \rangle \leq 1$, we have $\langle b, q^{(n)}(T) \rangle \to 0$ with probability 1. Convergence with probability 1, along with the bound $\langle b, q^{(n)}(T) - q^{(n)}(0) \rangle < C(b_1 + b_2)(T + 1)$ (recall the assumption of bounded arrivals), implies convergence in the mean too, and therefore,

$$\limsup_{n \to \infty} \sup_{\sum_{i \in I} b_i q_i^{(n)}(0) \leq 1} E\left[\sum_{i \in I} b_i q_i^{(n)}(T)\right] = 0. \quad (43)$$

By [24] (or see Theorem 4.1 of [4],) this proves the throughput-optimality of p-Log rule.

Fix a time $T > 0$ and constants $C^* > \delta > 0$. Just as (43) was obtained from (42), the following too can be shown. For all large $n$, uniformly on $\langle b, q^{(n)}(0) \rangle \leq C^*$, if $\langle b, q^{(n)}(\cdot) \rangle > \delta$ over $[t, t+T]$, then,

$$\sup_{\langle b, q^{(n)}(0) \rangle \leq C^*} E\left[\langle b, q^{(n)}(t+T) \rangle - \langle b, q^{(n)}(t) \rangle\right] \leq -\frac{\delta_5}{2}T.$$

The above along with Dynkin's formula (see Theorem 19.1.2 of [25]) implies the following result.

*Lemma 5:* Let constants $C^* > \delta > 0$ be fixed. Consider the stopping time,

$$\boldsymbol{\beta}^{(n)} = \inf\left\{t \geq 0: \langle b, q^{(n)}(t) \rangle \leq \delta\right\}.$$

Then, for all sufficiently large $n$, uniformly on the initial states with $\langle b, q^{(n)}(0) \rangle \leq C^*$, we have,

$$E[\boldsymbol{\beta}^{(n)}] \leq \Delta_2 C^*,$$

for some finite $\Delta_2 > 0$.

*Proof of Theorem 2:* Now we can proceed with the proof of Theorem 2. Consider the scaled (random) process $q^{(n)}$. For any fixed constants $C^* > 1 > \epsilon^* > \epsilon > \delta > 0$, define the following stopping times,

$$\boldsymbol{\alpha}^{(n)} = \inf\left\{t > 0: \langle b, q^{(n)}(t) \rangle \geq 1\right\},$$
$$\boldsymbol{\beta}^{(n)} = \inf\left\{t > 0: \langle b, q^{(n)}(t) \rangle \leq \delta\right\},$$
$$\boldsymbol{\eta}^{(n)} = \inf\left\{t > \boldsymbol{\beta}^{(n)}: \langle b, q^{(n)}(t) \rangle \geq \epsilon\right\}.$$

Let $p^{(n)}$ denote the stationary distribution of process $q^{(n)}(t)$, $p_x^{(n)}$ its distribution conditional on $q^{(n)}(0) = x$, and $E_x$ expectation under $p_x^{(n)}$. Then, for any arbitrary $T > 0$, it is easy to show (see (8.15) [15]) the following upper bound on the probability of overflow,

$$p^{(n)}\left(\sum_{i \in I} b_i q_i^{(n)} > 1\right) \leq \frac{\sup_{y: \langle b, y \rangle \leq C^*} E_y[\boldsymbol{\beta}^{(n)}]}{\inf_{z: \langle b, z \rangle \geq \epsilon} E_z[\boldsymbol{\eta}^{(n)}]} \times$$
$$\sup_{x: \langle b, x \rangle \leq \epsilon^*} \left(p_x^{(n)}(\boldsymbol{\beta}^{(n)} \geq T) + p_x^{(n)}(\boldsymbol{\alpha}^{(n)} \leq T)\right). \quad (44)$$

Now, by uniform upper bound in Lemma 5, we have $\limsup_{n \to \infty} \sup_{y: \langle b, y \rangle \leq C^*} E_y[\boldsymbol{\beta}^{(n)}] \leq \Delta_2 C^*$; and by bounded arrivals, we have the lower bound $\liminf_{n \to \infty} \inf_{z: \langle b, z \rangle \geq \epsilon} E_z[\boldsymbol{\eta}^{(n)}] > 0$. Therefore, the terms that are going to decide the limit,

$$\limsup_{n \to \infty} \frac{1}{n} \log p^{(n)}\left(\sum_{i \in I} b_i q_i^{(n)} > 1\right),$$

are $p_x^{(n)}(\boldsymbol{\beta}^{(n)} \geq T)$ and $p_x^{(n)}(\boldsymbol{\alpha}^{(n)} \leq T)$.

By Theorem 5, we can choose $T$ large enough such that $K(\epsilon^*, \delta, T) \geq K(C^*, \delta, T) \geq J_{**}$. Then by Theorem 4,

$$\limsup_{n \to \infty} \frac{1}{n} \log \sup_{x: \langle b, x \rangle \leq \epsilon^*} p_x^{(n)}(\boldsymbol{\beta}^{(n)} \geq T)$$
$$\leq -K(C^*, \delta, T) \leq -J_{**}.$$

By (41) and the definition of $J_{**}$ in (15), for any $\epsilon_6 > 0$, we can choose an even larger a $T$ (if required) and an $\epsilon^* > 0$ small enough such that $J_{**,\leq T,\epsilon^*} > J_{**} - \epsilon_6$. Then by Theorem 3,

$$\limsup_{n\to\infty} \frac{1}{n} \log \sup_{x:\langle b,x\rangle \leq \epsilon^*} p_x^{(n)}\left(\boldsymbol{\alpha}^{(n)} \leq T\right)$$
$$\leq -J_{**,\leq T,\epsilon^*} < -(J_{**} - \epsilon_6).$$

Since we can choose $\epsilon_6$ arbitrarily small (and subsequently fix constants $1 > \epsilon^* > \epsilon > \delta > 0$), substituting the above two bounds in (44) completes the proof. ∎

## Acknowledgment

The authors would like to thank A. Stolyar for providing an early copy of [15].